\documentclass[12pt]{article}
\newcounter{lyter}[equation]
\newcommand{\ictimes}{\mbox{$\times \hspace{-1em}\supset$}}
\title{Basic Theorem, Gauge Algebra,
$\theta$-superfield QED \protect \\ in the Lagrangian Formulation of
\protect \\
General Superfield Theory of Fields}
\author{A.A. Reshetnyak\thanks{E-mail: reshetnyak@ssti.ru}}
\date{\it Department of Mathematics, Seversk State Technological Institute,
\protect \\
Seversk, {\rm 636036}, Russia}
\begin{document}
\maketitle
\begin{abstract}
The  basic theorem of the Lagrangian formulation for general superfield
theory of fields (GSTF) is proved. The gauge transformations of general type
(GTGT) and gauge algebra of generators of GTGT (GGTGT) as the
consequences of the above theorem are studied.

It is established the gauge  algebra of GGTGT contains the one of
generators of gauge transformations of special type (GGTST) as one's
subalgebra. In the framework of Lagrangian formulation for GSTF the
nontrivial superfield model generalizing the model of Quantum Electrodynamics
and belonging to the class of gauge theory of general type (GThGT) with
Abelian gauge algebra of GGTGT is constructed.
\end{abstract}

\noindent
{\large PACS codes:  11.10.Ef, 11.15.-q, 12.20.-m, 03.50.-z \protect }
\section{Introduction}
\renewcommand{\theequation}{\arabic{section}.\arabic{equation}}
\renewcommand{\thelyter}{\alph{lyter}}

The Lagrangian and Hamiltonian formulations for GSTF of the superfield (with
respect to odd time $\theta$) models description, suggested in the papers
[1,2], had  permitted to solve the problem of constructing the
superfield Lagrangian (in usual sense) quantization method for general gauge
theories in the framework of  general superfield quantization method
(GSQM) in the Lagrangian formalism [3].

GSQM permits by means of the path integral method to quantize the ordinary
gauge models of the quantum field theory extended, in a natural way, to the
superfield GSTF models. GSQM contains the BV quantization method for gauge
theories [4] as the particular case under special choice of the generating
equation [3]. By the main resulting GSQM feature it appears the Ward
identities form for generating functionals (superfunctions) of Green's
functions, including the effective action, which reflect the fact of these
superfunctions invariance under their translation with respect to variable
$\theta$ along integral curve of solvable [1,2] Hamiltonian system constructed
with respect to quantum gauge fixed action superfunction $S^{\Psi}_{H}(\theta,
\hbar)$. It is the above Hamiltonian system with which the standard BRST
symmetry transformations are associated [3] under corresponding notations
providing the $\theta$-superfield realization of that symmetry.

Theorem 1, formulated in Ref.[1], on reduction of the 1st order with
respect to differentiation on $\theta $ system of $N$ ordinary differential
equations (ODE) to generalized normal form (GNF) in the case of its linear
(functional) dependence appears by the key one in GSTF [1,2] and
in GSQM [3] construction on the whole. The paper is devoted to its
proof, to the investigation of the GTGT and a gauge algebra of the GGTGT, to
the connection of the latters with GTST and a gauge algebra of the GGTST, to
the demonstration of the efficiency of these results on the example of
superfield (on $\theta$) quantum electrodynamics model.

In work the definitions, conventions and notations introduced in
Ref.[1] are made use unless otherwise stated.
\section{Proof of the Basic Theorem}

Consider the 2nd order with respect to derivatives on $\theta$ system of $N$
ODE in  normal form (NF)
\begin{eqnarray}
{\stackrel{\;\circ\circ}{g}\,}{}^i(\theta) = f^i\bigl(g(\theta),
{\stackrel{\;\circ}{g}}(\theta), \theta\bigr)\;,\; f^i(\theta) \in
C^{1}\bigl( T_{odd}{\cal N} \times \{\theta\} \bigr)\,,
\end{eqnarray}
in a some domain of the supermanifold ${\cal N}$ parametrized by local
coordinates $g^i(\theta), i = 1,\ldots,$ $N =(N_+, N_-)$ $(g^i(\theta)=g^i_0+
g^i_1\theta)$ being by unknown superfunctions\footnote{Because the index $i$
possesses by the complicated condensed contents then Eqs.(2.1) are, in
general, the system of partial differential equations [1]. The only
differential operator $\frac{d}{d\theta}$ is specially singled out here}.
Grassmann
parities $\varepsilon_P, \varepsilon_{\bar{J}}, \varepsilon$ of quantities
$g^i(\theta), g^i_0, g^i_1, f^i(\theta)$ are given by the formula ($
\varepsilon = \varepsilon_P + \varepsilon_{\bar{J}}$ [1])
\begin{eqnarray}
{} & (\varepsilon_P, \varepsilon_{\bar{J}}, \varepsilon)b(\theta) =
(\varepsilon_P(g^i_1)+1, \varepsilon_{\bar{J}}(g^i_1), \varepsilon(g^i_1)+1)=
(\varepsilon_P(g^i(\theta)), \varepsilon_{\bar{J}}(g^i(\theta)),
\varepsilon(g^i(\theta)))\,,{} &
\nonumber \\
{} & b(\theta)\in \{g^i(\theta), g^i_0,  f^i(\theta)\}\,. {} &
\end{eqnarray}
Eqs.(2.1) are equivalent to the following system of $2N$ ODE at most of
the 2nd order with respect to $\theta$
\begin{eqnarray}
{} & {} &{\stackrel{\;\circ\circ}{g}\,}{}^i(\theta) \equiv
\displaystyle\frac{d^2 g^i(\theta)}{d\theta^2\phantom{xxx}} \equiv
\displaystyle\frac{d\phantom{}}{d\theta}
\displaystyle\frac{d g^i(\theta)}{d\theta\phantom{xxx}}=0\,, \\ 
{} & {} & f^i\bigl(g(\theta), {\stackrel{\;\circ}{g}}(\theta),\theta\bigr)
=0\,,
\end{eqnarray}
so that the Cauchy problem setting for (2.1) is controlled by differential
constraints which are the subsystem of the 1st order on $\theta$ $N$ ODE
(2.4) [1].
In a general case the Eqs.(2.4) appear by (functionally) dependent.
Singling from (2.4) the independent subsystem of the 1st order on
$\theta$ ODE is effectively realized in fulfilling of the following
assumptions [1]:
\begin{eqnarray}
1)\hspace{0.2cm}\left(\tilde{g}{}^{i}(\theta),{\stackrel{\;\circ}{\tilde{g}}}{
}^i(\theta) \right) = (0,0) \in T_{odd}\Phi =
\left\{\left(g^{i}(\theta),{\stackrel{\;\circ}{g}}{}^i(\theta) \right)
\vert f^i\bigl(g(\theta), {\stackrel{\;\circ}{g}}(\theta),\theta\bigr) \equiv
0 \right\}\,, \hspace{2.1cm} 
\end{eqnarray}
2) $f^i\bigl(g(\theta), {\stackrel{\;\circ}{g}}(\theta),\theta\bigr)=0$
determines the  1st order smooth surface $T_{odd}\Phi$ in $T_{odd}{\cal N}$
which the condition holds on
\begin{eqnarray}
{\rm rank}_{\varepsilon_{\bar{J}}}
\left\|\frac{\delta_l f^{i}\bigl( g(\theta),
{\stackrel{\;\circ}{g}}(\theta), \theta\bigr)} {\delta
g^j(\theta_1)\phantom{xxxxxxxx}}\right\|_{\mid T_{odd} \Phi} \leq
N \equiv [f^{i}]\,.
\end{eqnarray}
Notion of the rank for supermatrix of the form (2.6) with respect to
$\varepsilon_{\bar{J}}$ grading for $f^i(\theta)$, $g^j(\theta_1)$ was
defined in [1,3] and
$\frac{\delta_l\phantom{xxx}}{\delta g^j(\theta_1)}$ denotes the left
superfield variational derivative with respect to superfunction
$g^j(\theta_1)$ ($\theta_1\ne \theta$).

\noindent
\underline{\bf Theorem 1}

\noindent
System of the 1st order on $\theta$ $N$ ODE with respect to $g^i(\theta)$
(2.4) subject to conditions (2.5), (2.6) being
unsolvable with respect to ${\stackrel{\;\circ}{g}}{}^i(\theta)$
is reduced to equivalent system of independent equations in GNF
under following nondegenerate parametrization for $g^i(\theta)=
(\alpha^{\bar{i}}(\theta),\beta^{\underline{i}}(\theta),\gamma^{
\sigma}(\theta))$, $i=(\bar{i},\underline{i},\sigma)$
\begin{eqnarray}
{\stackrel{\;\circ}{\alpha}}{}^{\bar{i}}(\theta)  =
\varphi^{\bar{i}}\bigl({\alpha}(\theta) , {\gamma}(\theta),
{\stackrel{\;\circ}{\gamma}}(\theta), \theta\bigr),\ \;
{\beta}^{\underline{i}}(\theta) =
\kappa^{\underline{i}}\bigl({\alpha}(\theta) , {\gamma}(\theta),
\theta\bigr)\;, 
\end{eqnarray}
with arbitrary superfunctions ${\gamma}^{\sigma}(\theta)$
and $\varphi^{\bar{i}}(\theta), \kappa^{\underline{i}}(\theta)$ $\in$ $
C^1(T_{odd}{\cal N}\times\{\theta\})$. The number
of $[{\gamma}^{\sigma}]$ coincides with one of differential identities among
Eqs.(2.4)
\begin{eqnarray}
\int d\theta f^{i}\bigl( g(\theta),
{\stackrel{\;\circ}{g}}(\theta), \theta\bigr) \check{{\cal R}}_{i
\sigma}\bigl( g(\theta), {\stackrel{\;\circ}{g}}(\theta), \theta;
{\theta}^{\prime}\bigr) = 0\;,
\end{eqnarray}
where operators
$\check{{\cal R}}_{i \sigma}\bigl( g(\theta),
{\stackrel{\;\circ}{g}}(\theta), \theta; {\theta}^{\prime}\bigr)$ are a)
local on $\theta$ and b) functionally independent ones
\begin{eqnarray}
{\rm a})\hspace{0.3cm} \check{{\cal R}}_{i \sigma}\bigl( g(\theta),
{\stackrel{\;\circ}{g}}(\theta), \theta; {\theta}^{\prime}\bigr) \equiv
\check{{\cal R}}_{i \sigma}(\theta; {\theta}^{\prime}) =
\sum_{k=0}^{1}\left(\left(\frac{d}{d\theta}\right)^{k}\delta(\theta -
\theta')\right) \check{{\cal R}}^{k}_{{}i \sigma}
\bigl(g(\theta),{\stackrel{\;\circ}{g}}(\theta), \theta\bigr)\,,
\hspace{0.9cm}
\end{eqnarray}
b) functional equation
\begin{eqnarray}
\int d{\theta}'\check{{\cal R}}_{i \sigma}(\theta;
{\theta}^{\prime})u^{\sigma}\bigl( g(\theta'),
{\stackrel{\;\circ}{g}}(\theta'), {\theta}^{\prime}\bigr) = 0,\
u^{\sigma}(\theta)\in C^1(T_{odd}{\cal N}\times\{\theta\}) 
\end{eqnarray}
has unique trivial solution.
\vspace{1ex}

\noindent
\underline{Proof} includes the investigation scheme of the corresponding
system of the 1st and 2nd orders with respect to even derivatives on $t\in{\bf
R}$ [5] because one can regard that $t\in i$ [1].

\noindent
{\bf 1)} In correspondence with (2.6) let us  assume that
\begin{eqnarray}
{\rm rank}_{\varepsilon_{\bar{J}}}
\left\|\frac{\delta_l f^{i}(\theta)}{\delta{\stackrel{\;\circ}{
g}}{}^j(\theta_1)\phantom{}}\right\|{\hspace{-0.5em}\phantom{\Bigr)}}_{
\mid T_{odd} \Phi}=
N-M \Longleftrightarrow
{\rm corank}_{\varepsilon_{\bar{J}}}
\left\|\frac{\delta_l f^{i}(\theta)}{\delta{\stackrel{\;\circ}{
g}}{}^j(\theta_1)\phantom{}}\right\|{\hspace{-0.5em}\phantom{\Bigr)}}_{
\mid T_{odd} \Phi}= M = (M_+,M_-)\,.
\end{eqnarray}
Then $f^i(\theta)$ as the
functions of ${\stackrel{\;\circ}{g}}{}^j(\theta)$ are dependent ones and
from (2.11) it follows  the possibility of the representation
\begin{eqnarray}
{} & {} & f^i(\theta) = ({\cal P}^a_1(\theta), p^A_1(\theta)),\;
a=1,\ldots,M; A=M+1,\ldots,N\,,\\
{} & {} & \hspace{-1em}
{\rm rank}_{\varepsilon_{\bar{J}}} \left\|
\displaystyle\frac{\partial_l p^{A}_1\bigl(g(\theta),{\stackrel{\;\circ}{
g}}(\theta), \theta\bigr)}{\partial
{\stackrel{\;\circ}{g}}{}^j(\theta)\phantom{xxxxxxxxx}}\right\|{\hspace{-0.5em}
\phantom{\Bigr)}}_{\mid T_{odd}\Phi}= N-M \Longleftrightarrow
{\rm corank}_{\varepsilon_{\bar{J}}}
\left\|\displaystyle\frac{\partial_l p^{A}_1(\theta)}{\partial
{\stackrel{\;\circ}{g}}{}^j(\theta)\phantom{x}}\right\|{\hspace{-0.5em}
\phantom{\Bigr)}}_{\mid T_{odd} \Phi}= M\,,\\
{} & {} & {\cal P}^a_1\bigl(g(\theta),{\stackrel{\;\circ}{g}}(\theta),
\theta\bigr) = p^{A}_1\bigl(g(\theta),{\stackrel{\;\circ}{g}}(\theta),
\theta\bigr)
\alpha_1^a{}_A\bigl(g(\theta),{\stackrel{\;\circ}{g}}(\theta), \theta\bigr) +
\Delta^a(g(\theta),\theta)\,. 
\end{eqnarray}
The superfunctions $\Delta^a(\theta)$ may be dependent ones, i.e.
\begin{eqnarray}
{\rm rank}_{\varepsilon_{\bar{J}}}\left\|\frac{\partial_l
\Delta^{a}(g(\theta),\theta)}{\partial
{g}^j(\theta)\phantom{xxxxx}}\right\|{\hspace{-0.5em}\phantom{\Bigr)}}_{
\mid\Phi}= M-K \leq M,\ 0\leq K=(K_+, K_-)
\,.
\end{eqnarray}
It means the superfunctions $\delta^{(1){}a_1}(g(\theta),\theta)$, $a_1=K+1,
\ldots,M$ exist that the condition holds
\begin{eqnarray}
{\rm rank}_{\varepsilon_{\bar{J}}}
\left\|\frac{\partial_l \delta^{(1){}a_1}(\theta)}{\partial
{g}^j(\theta)\phantom{xxx}}\right\|{\hspace{-0.5em}\phantom{\Bigr)}}_{
\mid\Phi}= M-K\,.
\end{eqnarray}
In (2.11), (2.13)--(2.16) the left partial superfield derivatives with
respect to
${\stackrel{\;\circ}{g}}{}^j(\theta), g^j(\theta)$ are denoted as
$\frac{\partial_l\phantom{xxx}}{\partial{\stackrel{\;\circ}{g}}{}^j(
\theta)}$,
$\frac{\partial_l\phantom{xxx}}{\partial g^j(\theta)}$ respectively [1].
By virtue of the assumption (2.5) for $\Delta^a(\theta)$,
${\cal P}^a_1(\theta)$ the following representation is valid
\begin{eqnarray}
{} & {} & \Delta^{a}(g(\theta),\theta) =
\delta^{(1){}a_1}(g(\theta),\theta)\beta_1^a{}_{a_1}(g(\theta),\theta),\
{\rm rank}_{\varepsilon_{\bar{J}}}
\left\|\beta_1^a{}_{a_1}(\theta)\right\|_{\mid\Phi}= M-K\,,
\nonumber \\
{} & {} &
{\cal P}^a_1\bigl(g(\theta),{\stackrel{\;\circ}{g}}(\theta), \theta\bigr) =
p^{A}_1(\theta)
\alpha_1^a{}_A(\theta) + \delta^{(1){}a_1}(\theta)\beta_1^a{}_{a_1}(\theta)
\,.
\end{eqnarray}
Divide the all ${\cal P}^a_1(\theta)$ onto
2 groups:  ${\cal P}^a_1(\theta)$ = $(\overline{\cal P}{}^{A_1}_1(\theta),
\underline{\cal P}{}^{a_1}_1(\theta))$, $A_{1}=1,\ldots,K$; $a_{1}=K+1,
\ldots, M, a=(A_{1}, a_{1})$
\begin{eqnarray}
{} \hspace{-0.5em}\underline{\cal P}{}^{a_1}_1\bigl(g(\theta),{\stackrel{
\;\circ}{g}}(\theta),
\theta\bigr)\hspace{-0.1em}=\hspace{-0.1em} p_1^A(\theta)
\underline{\alpha}{}^{a_1}_1{}_A\bigl(g(\theta),{\stackrel{\;\circ}{g}}(
\theta), \theta\bigr)\hspace{-0.1em}+ \hspace{-0.1em}
\delta^{(1){}b_1}(\theta)\underline{\beta}{}_1^{a_1
}{}_{b_1}(g(\theta),\theta),\, {\rm sdet}\left\|\underline{\beta}{}_1^{a_1
}{}_{b_1}(\theta)\right\| \neq 0. 
\end{eqnarray}
\begin{sloppypar}
Then from  (2.11)--(2.18) it follows that $f^i(\theta)=
(\overline{\cal P}{
}^{A_1}_1(\theta)$, $\underline{\cal P}{}^{a_1}_1(\theta)$, $p^A_1(\theta))$
are connected with $f^{j}_{(1)}\bigl(g(\theta), {\stackrel{\;\circ}{g
}}(\theta), \theta\bigr)$ =
$(\overline{\cal P}{
}^{B_1}_1(\theta)$, $\delta^{(1)b_1}(\theta), p^B_1(\theta))$ by means of
the
nondegenerate supermatrix $K^{0,1}(\theta)$ = $\left\|K^{0,1}{}^i{}_j\bigl(
g(\theta),
{\stackrel{\;\circ}{g}}(\theta),\theta\bigr)\right\|$ in $T_{odd}V$
$\supset$ $T_{odd}\Phi$ for some $V\subset{\cal N}$
\end{sloppypar}
\begin{eqnarray}
f^i(\theta) = f^{j}_{(1)}(\theta)K^{0,1}{}^i{}_j(\theta),\
\bigl(K^{0,1}(\theta)\bigr)^{-1} = K^{1,0}(\theta)\,,
\end{eqnarray}
providing the equivalence of Eqs.(2.4) and $f^{j}_{(1)}(\theta)=0$.
Consider the superfunctions $\overline{\cal P}{}^{A_1}_1(\theta)$ among $
{\cal P}^{a}_1(\theta)$ in (2.17). They are the identities for
superfunctions $f^{j}_{(1)}(\theta)$ or $f^i(\theta)$ and can be written
by means of two equivalent expressions
\begin{eqnarray}
\int d\theta f^{j}_{(1)}\bigl(g(\theta), {\stackrel{\;\circ}{g}}(\theta),
\theta\bigr){\cal R}^{(1)}_{jA_1}\bigl(g(\theta), {
\stackrel{\;\circ}{g}}(\theta), \theta;\theta'\bigr) = 0\,;
\end{eqnarray}
\vspace{-3ex}
\renewcommand{\theequation}{\arabic{section}.\arabic{equation}\alph{lyter}}
\begin{eqnarray}
\setcounter{lyter}{1}
\overline{\cal P}{}^{A_1}_1(\theta) & = &
\int d\theta' \underline{f}_{(1)}^{(a_1,A)}\bigl(g(\theta'), {\stackrel{\;
\circ}{g}}(\theta'),
\theta'\bigr)\Lambda_{(1){}(a_1,A)}{}^{A_1}\bigl(g(\theta'),
{\stackrel{\;\circ}{g}}(\theta'), \theta';\theta\bigr)
\,,\\
\setcounter{equation}{21}
\setcounter{lyter}{2}
f^{j}_{(1)}(\theta) & \equiv &
\left(\overline{\cal P}{}^{B_1}_1(\theta),
\underline{f}_{(1)}^{(b_1,B)}\bigl(g(\theta), {\stackrel{\;\circ}{g}}(\theta),
\theta\bigr)\right) \,.
\end{eqnarray}
Therefore,  among $f^{i}_{(1)}(\theta)=0$
the only $\underline{f}{}^{(a_1,A)}_{(1)}(\theta)=0$ are essential. It is the
latter equations are equivalent to (2.4).

\noindent
{\bf 2)} In  its turn the superfunctions ${\stackrel{\;\circ}{\delta}}{}^{
(1){}a_1}(\theta)$, $p_1^A(\theta)$ may be dependent ones with respect to
${\stackrel{\;\circ}{g}}{}^j(\theta)$
\renewcommand{\theequation}{\arabic{section}.\arabic{equation}}
\begin{eqnarray}
{} & {\rm rank}_{\varepsilon_{\bar{J}}}\left\|
\displaystyle\frac{\partial_l
\bigl({\stackrel{\;\circ}{\delta}}{}^{(1){}a_1}(\theta),
p_1^A (\theta)\bigr)}{\partial
{\stackrel{\;\circ}{g}}{}^j(\theta)\phantom{xxxxxxxxx}}\right\|{
\hspace{-0.5em}\phantom{\Bigr)}}_{\mid
T_{odd}\Phi}\hspace{-0.5em}= N-K -K_1 <[\delta^{(1)}(\theta)] +
[p_1(\theta)]\,, {} & \nonumber \\
{} &  N-K-K_1 \geq [\delta^{(1)}(\theta)], K_1=(K_{1+}, K_{1-})\,.{} &
\end{eqnarray}
It  follows from (2.22) the representability of $p_1^A(\theta)$ in the form
\begin{eqnarray}
{} & p_1^A(\theta)=
\bigl({\cal P}_2^{a_{11}}, p_2^{A_{11}}\bigr)(\theta),
a_{11} = M+1,\ldots,M+K_1; A_{11} = M+K_1+1,\ldots,N\,, {} &
\\
{} & {\cal P}_2^{a_{11}}\bigl(g(\theta),{\stackrel{\;\circ}{g}}(\theta),
\theta\bigr) = p_2^{A_{11}}
\bigl(g(\theta),{\stackrel{\;\circ}{g}}(\theta), \theta\bigr)
{\alpha_2}^{a_{11}}{}_{A_{11}}
\bigl(g(\theta),{\stackrel{\;\circ}{g}}(\theta), \theta\bigr)  +
{} & \nonumber \\
{} & \hspace{-1.0em} {\stackrel{\;\circ}{\delta}}{}^{(1){}a_1}(\theta)
{\nu_1}^{a_{11}}{}_{a_1}
\bigl(g(\theta),{\stackrel{\;\circ}{g}}(\theta), \theta\bigr) +
 \delta^{(1){}a_1}(\theta){\mu_1}^{a_{11}}{}_{a_1}
\bigl(g(\theta), \theta\bigr) +
\delta^{(2){}a_2}(g(\theta), \theta){\beta_2}^{a_{11}}{}_{a_2}
\bigl(g(\theta),\theta\bigr),{} & \\
{} & {\rm rank}_{\varepsilon_{\bar{J}}}
\left\|{\beta_2}^{a_{11}}{}_{a_2}(\theta)\right\|=[\delta^{(2)}(
\theta)] = M_1=(M_{1+},M_{1-})\,,{} & \nonumber \\
{} & a_2= M+K_1 - M_1 +1,\ldots,M+K_1\,,{} &\\
{} &
{\rm rank}_{\varepsilon_{\bar{J}}} \left\|\displaystyle\frac{
\partial_l \bigl({\stackrel{\;\circ}{\delta}}{}^{(1){}a_1}(\theta),
p_2^{A_{11}}(\theta)\bigr)}{\partial
{\stackrel{\;\circ}{g}}{}^j(\theta)\phantom{xxxxxxxxx}}\right\|{
\hspace{-0.5em}\phantom{\Bigr)}}_{\mid
T_{odd}\Phi}=  [\delta^{(1)}(\theta)]+ [p_2(\theta)] = N-K-K_1\,,{} &
\end{eqnarray}
where $\delta^{(2)a_2}(\theta)$ are the
superfunctions being independent ones on $\delta^{(1){}a_1}(\theta)$
\begin{eqnarray}
{\rm rank}_{\varepsilon_{\bar{J}}} \left\|
\frac{\partial_l \bigl({\delta}^{(1){}a_1}(\theta),
\delta^{(2){}a_{2}}(\theta)\bigr)}{\partial
{g}^j(\theta)\phantom{xxxxxxxxxxx}}\right\|{\hspace{-0.5em}\phantom{\Bigr)}}_{
\mid\Phi}=  [\delta^{(1)}(
\theta)]+ [\delta^{(2)}(\theta)] = M+M_1-K\,.
\end{eqnarray}
According to (2.24)--(2.27) divide  ${\cal P}_2^{a_{11}}(\theta)$ onto 2
groups
\begin{eqnarray}
 {} & \hspace{-1.0em}{\cal P}_2^{a_{11}}(\theta) =
\bigl(\overline{\cal P}{}_2^{A_2},
\underline{\cal P}{}_2^{a_2}\bigr)\bigl(g(\theta),{\stackrel{\;\circ}{g}}(
\theta), \theta\bigr),
A_2 = M+1,\ldots,M+K_1-M_1, {} &\\
{} & \underline{\cal P}{}_2^{a_2}(\theta) =
p_2^{A_{11}}(\theta)
{\underline{\alpha}{}_2}^{a_2}{}_{A_{11}}(\theta) +
{\stackrel{\;\circ}{\delta}}{}^{(1){}a_1}(\theta)
{\underline{\nu}{}_1}^{a_2}{}_{a_1}(\theta)+ {} & \nonumber \\
{} & {\delta}^{(1){}a_1}(\theta){\underline{\mu}{}_1}^{a_2}{}_{a_1}(\theta)+
\delta^{(2){}b_2}(\theta){\underline{\beta}{}_2}^{a_2}{}_{b_2}(\theta),\;
{\rm sdet}\left\|{\underline{\beta}{}_2}^{a_2}{}_{b_2}(\theta\right\|\ne 0
\,.{} & 
\end{eqnarray}
and define the set of superfunctions $f^{j}_{(2)}\bigl(g(\theta),{\stackrel{
\;\circ}{g}}(\theta),\theta\bigr)$ = $(\overline{\cal P}{}^{A_1}_1(\theta)$,
$\overline{\cal P}{}^{A_2}_2(\theta)$, $\delta^{(1)a_1}(\theta)$, $\delta^{(2)
a_2}(\theta)$, $p^{A_{11}}_2(\theta))$
connected with $f^{i}_{(1)}\bigl(g(\theta),{\stackrel{\;\circ}{g}}(\theta),
\theta\bigr)$ =
$(\overline{\cal P}{
}^{B_1}_1(\theta)$, $\delta^{(1)b_1}(\theta)$, $\overline{\cal P}{}^{B_2}_2(
\theta)$, $\underline{\cal P}{}^{b_2}_2(\theta)$, $p^{B_{11}}_2(\theta))$
(and therefore with $f^j(\theta)$ (2.4)) by the
nondegenerate supermatrix $K^{1,2}(\theta)$ = $\left\|K^{1,2}{}^i{}_j
\bigl(g(\theta),
{\stackrel{\;\circ}{g}}(\theta),\theta\bigr)\right\|$ in $T_{odd}V$
\begin{eqnarray}
{} & {} & f^{i}_{(1)}(\theta) = f^{j}_{(2)}(\theta)K^{1,2}{}^i{}_j(\theta),\
K^{1,2}(\theta) = \left\|
\begin{array}{ccccc}
\delta^{B_1}{}_{A_1} & 0 & 0 & 0 & 0 \\
0 & 0 & \delta^{B_2}{}_{A_2} & 0 & 0 \\
0 & \delta^{b_1}{}_{a_1} & 0 & A^{b_2}{}_{a_1}(\theta) & 0 \\
0 & 0 & 0 & B^{b_2}{}_{a_2}(\theta) & 0 \\
0 & 0 & 0 & C^{b_2}{}_{A_{11}}(\theta) & \delta^{B_{11}}{}_{A_{11}} \\
\end{array}\right\|,\nonumber \\
{} & {} & A^{b_2}{}_{a_1}(\theta) = \displaystyle\frac{\stackrel{
\leftarrow}{d}}{d\theta}\underline{\nu}{}_1{}^{b_2}{}_{a_1}(\theta) +
\underline{\mu}{}_1{}^{b_2}{}_{a_1}(\theta),
B^{b_2}{}_{a_2}(\theta)=\underline{\beta}{}_2{}^{b_2}{}_{a_2}(\theta),
C^{b_2}{}_{A_{11}}(\theta) = \underline{\alpha}{}_2{}^{b_2}{}_{A_{11}}(
\theta)\,.
\end{eqnarray}
Its inverse supermatrix has the form
\begin{eqnarray}
K^{2,1}(\theta) =(K^{1,2}(\theta))^{-1} =
\left\|
\begin{array}{ccccc}
\delta_{B_1}{}^{C_1} & 0 & 0 & 0 & 0 \\
0 & 0 & \delta_{b_1}{}^{c_1} & -(A B^{-1})_{b_1}{}^{c_2}(\theta) & 0 \\
0 & \delta_{B_2}{}^{C_2} & 0 & 0 & 0 \\
0 & 0 & 0 & (B^{-1})_{b_2}{}^{c_2}(\theta) & 0 \\
0 & 0 & 0 & - (C B^{-1})_{B_{11}}{}^{c_2}(\theta) & \delta_{B_{11}}{}^{C_{11}}
\\
\end{array}\right\|\,.
\end{eqnarray}
From (2.30), (2.31) it follows both $K^{1,2}(\theta)$ and $K^{2,1}(\theta)$
appear by the local differentiation operators with
respect to $\theta$. In its
turn from (2.22)--(2.24) it implies the additional, already differential on
$\theta$, identities exist among $f^{j}_{(2)}(\theta)$ $\equiv$ $(
\overline{\cal P}{}^{A_1}_1(\theta)$, $\underline{f}{}^{(a_1,A)}_{(1)}(
\theta))$ besides of ones (2.20)
\begin{eqnarray}
\int d\theta \underline{f}{}_{(1)}^{(a_1, A)}\bigl(g(\theta),
{\stackrel{\;\circ}{g}}(\theta),
\theta\bigr){\cal R}^{(2)}_{(a_1, A){}A_2}\bigl(g(\theta),
{\stackrel{\;\circ}{g}}(\theta), \theta;\theta'\bigr) = 0
\,.
\end{eqnarray}
One can equivalently represent them in the form
\renewcommand{\theequation}{\arabic{section}.\arabic{equation}\alph{lyter}}
\begin{eqnarray}
\setcounter{lyter}{1}
\overline{\cal P}{}^{A_2}_2(\theta) & = &
\int d\theta' \underline{f}_{(2)}^{(a_1,a_2, A_{11})}\bigl(g(\theta'),
{\stackrel{\;\circ}{g}}(\theta'), \theta'\bigr)
\Lambda_{(2){}(a_1,a_2,A_{11})}{}^{A_2}\bigl(g(\theta'),
{\stackrel{\;\circ}{g}}(\theta'), \theta';\theta\bigr)
\,,\\
\setcounter{equation}{33}
\setcounter{lyter}{2}
f^{j}_{(2)}(\theta) & \equiv &
\left(\overline{\cal P}{}^{A_1}_1(\theta),\overline{\cal P}{}^{A_2}_2(\theta),
\underline{f}{}_{(2)}^{(a_1,a_2,A_{11})}(\theta)\right)\,.
\end{eqnarray}
Thus the only $\underline{f}{}^{(a_1,a_2,A_{11})}_{(2)}(\theta)=0$ are
the essential equations from $\underline{f}^{(a_1,A)}_{(1)}(\theta)=0$. It is
the former equations are completely equivalent to the initial ones (2.4).

\noindent
{\bf 3)} Let us assume the set of superfunctions $p_2^{A_{11}}(\theta)$,
${\stackrel{\;\circ}{\delta}}{}^{(1)a_1}(\theta)$, ${\stackrel{\;\circ}{
\delta}}{}^{(2)a_2}(\theta)$  are dependent with  respect to variables
${\stackrel{\;\circ}{g}}{}^j(\theta)$. Then, analogously
to above case the new constraints $\delta^{(3)}(g(\theta),\theta)$
arise being by independent on $\delta^{(1)}(\theta), \delta^{(2)}(\theta)$.
In view of the finiteness of the discrete part of index $i$ and in ignoring of
the covariance requirement with respect to $i$ it follows from
induction principle the existence of the $l$th step ($l\leq N$)
of the iterative procedure the such that the equations
\renewcommand{\theequation}{\arabic{section}.\arabic{equation}}
\begin{eqnarray}
{} &\underline{f}{}_{(l)}^{((a)_l,A_{l-1{}1})}
\bigl(g(\theta),{\stackrel{\;\circ}{g}}(\theta), \theta\bigr)\equiv
\underline{f}{}_{(l)}^{(a_1,\ldots,a_l,A_{l-1{}1})}
\bigl(g(\theta),{\stackrel{\;\circ}{g}}(\theta), \theta\bigr)= 0\,, {} &
\nonumber \\
{} & \underline{f}{}_{(l)}^{((a)_l,A_{l-1{}1})}(\theta) =
\Bigl(\delta^{(1){}a_1}(g(\theta),\theta),\ldots,\delta^{(l){}a_l}(g(\theta),
\theta), p_l^{A_{l-1{}1}} \bigl(g(\theta),{\stackrel{\;\circ}{g}}(\theta),
\theta\bigr)\Bigr) {} &
\end{eqnarray}
are equivalent to Eqs.(2.4) and
\begin{eqnarray}
{} &\hspace{-1.1em}{\rm rank}_{\varepsilon_{\bar{J}}}\hspace{-0.3em}\left\|
\displaystyle\frac{\partial_l
\bigl({\stackrel{\;\circ}{\delta}}{}^{(1){}a_1}(\theta),
\ldots, {\stackrel{\;\circ}{\delta}}{}^{(l){}a_l}(\theta), p_l^{A_{l-1{}1}}
(\theta)\bigr)}{\partial
{\stackrel{\;\circ}{g}}{}^j(\theta)\phantom{xxxxxxxxxxxxxxxxxxx}}
\right\|{\hspace{-0.5em}\phantom{\Bigr)}}_{\mid
T_{odd}\Phi}\hspace{-1.2em}
=\hspace{-0.15em} N\hspace{-0.15em}-\hspace{-0.15em} K\hspace{-0.15em} +
\hspace{-0.15em}\sum\limits_{s=1}^{l-1}(M_{s}-K_s)\hspace{-0.15em} =
\hspace{-0.15em}[p_l]\hspace{-0.15em} + \hspace{-0.15em}\sum\limits_{
s=1}^{l}[\delta^{(s)}],{} &\\
{} & {\rm rank}_{\varepsilon_{\bar{J}}} \left\|
\displaystyle\frac{\partial_l \bigl({\delta}^{(1){}a_1}(\theta),\ldots,
\delta^{(l){}a_{l}}(\theta)\bigr)}{\partial
{g}^j(\theta)\phantom{xxxxxxxxxxxxxx}}\right\|{\hspace{-0.5em}\phantom{\Bigr)}
}_{\mid\Phi}=
\sum\limits_{s=1}^{l}[\delta^{(s)}] = M-K+\sum\limits_{s=1}^{l-1}M_{s}
\,,{} &\\
{} & \hspace{-1em}
[p_l] = N-M-\sum\limits^{l-1}_{k=1}K_k,
[\delta^{(s)}]=M_s, s=\overline{1,l}\,.{} & 
\end{eqnarray}
Formally, the constructed algorithm  of  system (2.4) reduction
to GNF can  be written as follows
\begin{eqnarray}
f^i(\theta) = f^{j}_{(1)}(\theta)K^{0,1}{}^i{}_j(\theta)\,,
\end{eqnarray}
\vspace{-3ex}
\renewcommand{\theequation}{\arabic{section}.\arabic{equation}\alph{lyter}}
\begin{eqnarray}
\setcounter{lyter}{1}
{} & {} & \hspace{-2.5em}
f^{i}_{(1)}(\theta) = \bigl(\overline{\cal P}{}_1^{A_1}(\theta),
\underline{f}{}_{(1)}^{(a_1,A)}(\theta)\bigr)\,, \\
\setcounter{equation}{39}
\setcounter{lyter}{2}
{} & {} &\hspace{-2.5em}
\overline{\cal P}{}^{A_1}_1(\theta)  =
\int d\theta' \underline{f}{}_{(1)}^{(a_1,A)}(\theta')
\Lambda_{(1){}(a_1,A)}{}^{A_1}(\theta';\theta)\Longleftrightarrow
\int d\theta f^{j}_{(1)}(\theta){\cal R}^{(1)}_{j{}A_1}(\theta;\theta') = 0\,;
\\
\setcounter{equation}{39}
\setcounter{lyter}{3}
{} & {} & \hspace{-3.0em}
 f^{i}_{(1)}(\theta) = f^{j}_{(2)}(\theta)K^{1,2}{}^i{}_j(\theta)
\,,\\
\setcounter{equation}{40}
\setcounter{lyter}{1}
{} & {} & \hspace{-3.0em}
f^{i}_{(2)}(\theta) = \bigl(\overline{\cal P}{}_1^{A_1}(\theta),
\overline{\cal P}{}_2^{A_2}(\theta),
\underline{f}{}_{(2)}^{(a_1,a_2,A_{11})}(\theta)\bigr)=
\bigl(\overline{\cal P}{}_1^{A_1}(\theta),
\underline{f}{}_{(1)}^{(a_1,A)}(\theta)\bigr)
\,,\\
\setcounter{equation}{40}
\setcounter{lyter}{2}
{} & {} & \hspace{-3.0em}
\overline{\cal P}{}_2^{A_2}(\theta)\hspace{-0.1em} =\hspace{-0.2em}
\int\hspace{-0.2em} d\theta'\hspace{-0.1em} \underline{f}{}_{(2)}^{((a)_2,A_{
11})}(\theta')
\Lambda_{(2){}((a)_2,A_{11})}{}^{A_2}(\theta';\theta)
\hspace{-0.2em}\Longleftrightarrow\hspace{-0.2em}
\int\hspace{-0.2em} d\theta\hspace{-0.1em}
\underline{f}{}_{(1)}^{(a_1,A)}(\theta){\cal R}^{(2)}_{(a_1,A_1){}A_2}(
\theta;\theta')\hspace{-0.1em} =\hspace{-0.1em} 0;\\
\setcounter{equation}{40}
\setcounter{lyter}{3}
{} & {} & \hspace{-3.0em}
f^{i}_{(2)}(\theta) = f^{j}_{(3)}(\theta)K^{2,3}{}^i{}_j(\theta)\,,
\end{eqnarray}
\vspace{-3ex}
\begin{eqnarray}
\ldots & \ldots &  \ldots \nonumber
\end{eqnarray}
\vspace{-3ex}
\begin{eqnarray}
\setcounter{equation}{41}
\setcounter{lyter}{1}
{} & {} & \hspace{-3.5em}
f^{i}_{(l)}(\theta) =
\bigl(\overline{\cal P}{}_1^{A_1}(\theta),
\ldots, \overline{\cal P}{}_{l}^{A_{l}}(\theta),
\underline{f}{}_{(l)}^{((a)_{l},A_{l-1{}1})}(\theta)\bigr)\hspace{-0.1em}=
\hspace{-0.1em}
\bigl(\overline{\cal P}{}_1^{A_1}(\theta),\ldots,\overline{\cal P}{}_{l-1}^{
A_{l-1}}(\theta),\underline{f}{}_{(l-1)}^{((a)_{l-1},A_{l-2{}1})}(
\theta)\bigr),\\
\setcounter{equation}{41}
\setcounter{lyter}{2}
{} & {} &\hspace{-3.0em}
\overline{\cal P}{}_l^{A_l}(\theta) =
\int d\theta' \underline{f}{}_{(l)}^{((a)_l,A_{l-1{}1})}(\theta')
\Lambda_{(l){}((a)_l,A_{l-1{}1})}{}^{A_l}(\theta';\theta)
\Longleftrightarrow \nonumber \\
{} & {} & \hspace{-0.5em} \int d\theta
\underline{f}{}_{(l-1)}^{((a)_{l-1},A_{l-2{}1})}(\theta){\cal
R}^{(l)}_{((a)_{l-1},A_{l-2{}1}){}A_l}(\theta;\theta') = 0\,.
\end{eqnarray}
Thus, the Eqs.(2.4) have been reduced to equivalent ones in GNF (2.34)
being by functionally independent. Comparison of (2.34) with
(2.7) means by virtue of (2.35), (2.36) that
\renewcommand{\theequation}{\arabic{section}.\arabic{equation}}
\begin{eqnarray}
{} & {\stackrel{\;\circ}{\alpha}}{}^{\bar{i}}(\theta)  -
\varphi^{\bar{i}}\bigl({\alpha}(\theta) , {\gamma}(\theta),
{\stackrel{\;\circ}{\gamma}}(\theta), \theta\bigr)=0\Longleftrightarrow
p_l^{A_{l-1{}1}}
\bigl(g(\theta),{\stackrel{\;\circ}{g}}(\theta), \theta\bigr)=0,\
\bar{i}=A_{l-1{}1}\,,{} &\\
{} & {\beta}^{\underline{i}}(\theta) -
\kappa^{\underline{i}}\bigl({\alpha}(\theta) , {\gamma}(\theta),
\theta\bigr)=0\Longleftrightarrow
\delta^{(k){}a_k}(g(\theta),\theta)=0,\ k=1,\ldots,l, \underline{i}=(a_1,
\ldots,a_l)
\,.{} & 
\end{eqnarray}
The  number of arbitrary superfunctions $\gamma(\theta)$ in (2.7) is equal to
\begin{eqnarray}
[\gamma(\theta)] = [g^i(\theta)]-[\alpha^{\bar{i}}(\theta)]-
[{\beta}^{\underline{i}}(\theta)]=[f^i(\theta)] - [\underline{f}{}^{((a)_l,
A_{l-1{}1})}_{(l)}(\theta)]=K+ \sum\limits^{l-1}_{s=1}(K_s - M_s)
\end{eqnarray}
and coincides with  one of the identities in (2.39b), (2.40b)$,\ldots,$
(2.41b).

As far as the supermatrices $K^{s,s-1}(\theta)$ = $(K^{s-1,s}(\theta))^{-1}$
are the local ones on $\theta$, $s=1,\ldots,l$, then one can write the
identities in terms of the initial equations (2.4) which have the form (2.8)
with local on $\theta$ and functionally independent operators
$\check{\cal R}_{i\sigma}(\theta;\theta')$  being by polynomials with
respect to $K^{s,s-1}(\theta)$.

\vspace{1ex}
\noindent
\underline{\bf Remark}: The above proof have not concerned the possible
complicated structure of index $i$ (see footnote 1). The locality of operators
$\check{\cal R}_{i\sigma}(\theta;\theta')$  with respect to other
continual parts of the indices $i, \sigma$ may be shown in the
analogous way. However, the requirement of functional independence of
$\check{\cal R}_{i\sigma}(\theta;\theta')$ leads, in general case, to
the loss of covariance for these quantities.

From the Theorem 1 proof  the validity  of its consequence [1] easily
follows (with use of the integer-valued functions of degree and least
degree:  ${\rm deg}_{c(\theta)}$, ${\rm min\,deg}_{c(\theta)}$, $c(\theta)\in$
$\{g^i(\theta)$, ${\stackrel{\;\circ}{g}}{}^i(\theta)$, $g^i(\theta){
\stackrel{\;\circ}{g}}{}^j(\theta),\ldots\}$ [1]).

\noindent
\underline{\bf Corollary 1}
\nopagebreak

If $f^i(\theta)$ (2.4) are the holonomic constraints
\begin{eqnarray}
{\rm deg}_{{\stackrel{\;\circ}{g}}(\theta)}f^i(\theta) =0\,,
\end{eqnarray}
then for $f^i(g(\theta),\theta)$ under
following parametrization of  superfunctions $g^i(\theta)$ $\mapsto$
$g'{}^i(g(\theta))$
\begin{eqnarray}
g'{}^i(\theta) = (\alpha^A(\theta),\gamma^{\sigma}(\theta)),\ i=(A, \sigma),
\sigma=1,\ldots,[\gamma], A=1,\ldots,[\alpha]
\end{eqnarray}
there exists the equivalent system of the holonomic constraints
\begin{eqnarray}
\Phi^A(\alpha(\theta),\gamma(\theta),\theta)=0\,.
\end{eqnarray}
The number $[\gamma]$ coincides  with one of algebraic (in the sense of
differentiation with respect to  $\theta$) identities among $f^i(\theta)$
\begin{eqnarray}
f^i(g(\theta),\theta)\check{\cal R}{}^{(0)}_{i\sigma}(g(\theta),\theta)=0
\end{eqnarray}
being obtained from (2.8) by means of integration on $\theta$ with allowance
made for validity of the type (2.9) connection  of $\check{\cal R}_{i\sigma}(
\theta,\theta')$ with algebraic (on $\theta$) operators
$\check{\cal R}{}^{(0)}_{i\sigma}(\theta)$
\begin{eqnarray}
\check{\cal R}_{i\sigma}(g(\theta),\theta;\theta') = \delta(\theta-\theta')
\check{\cal R}{}^{(0)}_{i\sigma}(g(\theta),\theta)(-1)^{\varepsilon(g^i(
\theta))}\,.
\end{eqnarray}
A dependence upon ${\stackrel{\;\circ}{g}}{}^i(\theta)$ in
(2.49) may be only parametric one.
\section{Application to  GSTF in Lagrangian Formulation}
\setcounter{equation}{0}

Consider as  ${\cal N}$ the supermanifold ${\cal M}_{cl}$ parametrized by
the classical superfields
${\cal A}^{\imath}(\theta)$
$$
{\cal A}^{\imath}(\theta)=A^{\imath} + \lambda^{\imath}\theta,\
(\varepsilon_P, \varepsilon_{\bar{J}}, \varepsilon){\cal A}^{\imath}(\theta)=
((\varepsilon_P)_{\imath}, (\varepsilon_{\bar{J}})_{\imath},
\varepsilon_{\imath}),\ \imath=1,\ldots,n = (n_+,n_-)\,,
$$
being by superfunctions defined on  ${\cal M}=\tilde{\cal M}\times\tilde{P}$,
in its turn to be  the quotient space of the supergroup $J=\bar{J}\times P$ =
$(\overline{M}\ictimes\bar{J}_{\tilde{A}}) \times P$: ${\cal M}=J/{J}_{
\tilde{A}}$ with one-parametric subsupergroup $P$ generated by the
Grassmann nilpotent variable $\theta$ [1]. Superspace ${\cal M}$ may be
parametrized by sets of supernumbers $(z^a,\theta)=(x^{\mu},\theta^{Ak},
\theta)$, if the representation for $\tilde{\cal M}$ is valid [1]
\[
\tilde{\cal M} = {\bf R}^{1,D-1\vert L{}c},\ \mu=0,1,\ldots,D-1,\,
A=1,\ldots,c=2^{[D/2]},\, k=\overline{1,L}\,,
\]
meaning that $\tilde{\cal M}$ appears by the real $D$-dimensional Minkowski
superspace with $L$ supersymmetries (if $\bar{J}$ is the
corresponding group of the usual $L$-extended supersymmetry). Superfields
${\cal A}^{\imath}(\theta)$ are considered by belonging to the special
Berezin superalgebra $\tilde{\Lambda}_{D\vert Lc+1}(z^{a},\theta;{\bf K})$,
${\bf K}=({\bf R}$ or ${\bf C})$ [1--3].

The $\Lambda_1(\theta,{\bf R})$-valued superfunction $S_L(\theta)$ $\equiv$
$S_L\bigl({\cal A}(\theta), {\stackrel{\ \circ}{\cal A}}(\theta),\theta\bigr)$
$\in$
$C^k(T_{odd}{\cal M}_{cl}\times\{\theta\})$, $k\leq\infty$, $(\varepsilon_P$,
$\varepsilon_{\bar{J}}$, $\varepsilon)S_L(\theta)$ = $(0,0,0)$,
$T_{odd}{\cal M}_{cl}$ = $\bigl\{\bigl({\cal A}^{\imath}(\theta)$, ${
\stackrel{\ \circ}{\cal A}}{}^{\imath}(\theta)\bigr)\vert{\cal A}^{\imath}(
\theta)$ $\in$ ${\cal M}_{cl}\bigr\}$ and superfunctional $Z[{\cal A}]$ =
$\int d\theta S_L(\theta)$, $Z[{\cal A}]$ $\in$ $C_{F}$,
$(\varepsilon_P$, $\varepsilon_{\bar{J}}$, $\varepsilon)Z[{\cal A}]$ =
$(1,0,1)$  are
the central objects in the Lagrangian formulation for GSTF characterizing
the superfield model on this stage [1] of investigation. Dynamics of the model
follows from a variational principle for $Z[{\cal A}]$ and is
described by Euler-Lagrange equations [1]
\begin{eqnarray}
{\cal L}_{\imath}^{l}(\theta) S_{L}(\theta)\equiv
\left(\frac{\partial_l \phantom{xxx}}{
\partial{\cal A}^{\imath}(\theta)}
 -(-1)^{\varepsilon_{\imath}}\frac{d}{d\theta}\frac{\partial_l
\phantom{xxx}}{\partial{\stackrel{\ \circ}{\cal
A}}{}^{\imath}(\theta)}\right)S_{L}(\theta) =
\frac{\delta_l Z[{\cal A}]}{\delta{\cal A}^{\imath}(\theta)} = 0\,,
\end{eqnarray}
being represented equivalently by virtue of  (2.1), (2.3), (2.4) by the
Lagrangian system (LS)
\begin{eqnarray}
{} & {} &
{\stackrel{\,\circ\circ}{\cal A}}{}^{\jmath}(\theta) \displaystyle\frac{
\partial^{2}_{l} S_{L}(\theta) \phantom{xxxx}}
{{\partial{\stackrel{\ \circ}{\cal
A}}{}^{\imath}(\theta)}{\partial{\stackrel{\ \circ}{\cal A}}{}^{\jmath}
(\theta)}} = 0
\,,\\
{} & {} & {\Theta}_{\imath}\bigl( {{\cal A}}(\theta), {\stackrel{\
\circ}{{\cal A}}}(\theta), \theta \bigr) \equiv
\displaystyle\frac{
\partial_l S_{L}(\theta) }{\partial{\cal A}^{\imath}(\theta)}
 -(-1)^{\varepsilon_{\imath}}\left(\frac{\partial_l}{\partial\theta}\frac{
 \partial_l
S_{L}(\theta)}{{\partial{\stackrel{\ \circ}{\cal A}}{}^{\imath}(\theta)}} +
{\stackrel{\ \circ}{\cal A}}{}^{\jmath}(\theta)\displaystyle\frac{
\partial_l \phantom{xxx}}{\partial{\cal A}^{\jmath}(\theta)}
\displaystyle\frac{\partial_l S_{L}(\theta)}
{{\partial{\stackrel{\ \circ}{\cal A}}{}^{\imath}(\theta)}} \right) = 0
\,.
\end{eqnarray}
Subsystem (3.3), for ${\rm deg}_{{\stackrel{\ \circ}{\cal A}}(\theta)}\Theta_{
\imath}(\theta) \ne 0$ called the differential constraints in Lagrangian
formalism
(DCLF) (for ${\rm deg}_{{\stackrel{\ \circ}{\cal A}}(\theta)}\Theta_{\imath}(
\theta) = 0$  the
holonomic constraints in Lagrangian formalism (HCLF)), restricts an
arbitrariness in a choice of $2n$ initial conditions
$\left(\bar{{\cal A}}^{\imath}(0), \bar{\stackrel{
\ \circ}{\cal A}\;}{}^{\imath}(0)\right)$ for $\theta = 0$ in
setting of Cauchy problem. Subsystem (3.2) are not written in NF with respect
to ${\stackrel{\,\circ\circ}{\cal A}}{}^{\imath}(\theta)$ and
possibility to pass to NF depends on the  nondegeneracy of the supermatrix
$K(\theta)$ = $
\left\|\frac{\partial_{r}\phantom{xxx}}{{\partial{\stackrel{\ \circ}{\cal
A}}{}^{\jmath}(\theta)}} \frac{\partial_l
 S_{L}(\theta)}{{\partial{\stackrel{\ \circ}{\cal A}}{}^{\imath}(\theta)}}
 \right\|$.

DCLF themselves  appear, in general, by dependent system of the 1st order on
$\theta$ $n$ ODE with respect to unknowns ${\cal A}^{\imath}(\theta)$.
Reduction of $\Theta_{\imath}(\theta)$ to GNF is realized independently on
subsystem of the 2nd order on $\theta$ $n$ ODE (3.2) in the
result of Theorem 1 application directly to (3.3). To this end let us adapt
the assumption (2.5), (2.6) to the case of the Lagrangian GSTF [1,3] only
in terms of $Z[{\cal A}]$:
\begin{eqnarray}
{}\hspace{-4em} 1) {}\hspace{4em} \exists
\left({{\cal A}}^{\imath}_{0}(\theta),
{\stackrel{\circ}{{\cal A}^{\imath}_{0}}}(\theta)\right)
\in T_{odd}{\cal M}_{cl}:
{\Theta}_{\imath}(\theta)_{\Bigl| \left({\cal
A}(\theta), {\stackrel{\ \circ}{\cal A}}(\theta)\right) = \left( {\cal
A}_0(\theta), {\stackrel{\ \circ}{\cal A}}_0(\theta)\right)} = 0\,;
\end{eqnarray}
\vspace{-3ex}
\begin{eqnarray}
{} & {}\hspace{-2em}
2) {}\hspace{2em} \exists\Sigma \subset {\cal M}_{cl}\; \mbox{(
$\Sigma$--smooth supersurface)}: \left( {{\cal
A}}^{\imath}_{0}(\theta), {\stackrel{\ \circ}{\cal A}}{}^{\imath}_{0}(
\theta)\right) \in T_{odd}\Sigma,\ \Theta_{\imath}(\theta)_{\mid T_{
odd}\Sigma} = 0\,, {} &  \\ 
{} & {\rm dim}_{\varepsilon}{\Sigma}=m=(m_+,m_-), {\rm dim}_{\varepsilon
}{T_{odd}\Sigma}\equiv
{\rm dim}{T_{odd}\Sigma}= (m_+ +m_-,m_- +m_+)\,; {} &  
\end{eqnarray}
index $\imath$ can be divided $\imath=(A,\alpha)$, $A=1,\ldots,n-m$,
$\alpha=n-m+1,\ldots,n$ in a such way that the condition  holds
\begin{eqnarray}
{\rm rank}_{\varepsilon_{\bar{J}}} \left\| \frac{\delta_l \phantom {xxxx}}{
\delta{\cal A}^{\jmath}(\theta_1)} \frac{\delta_l Z[{\cal A}]}{\delta{\cal
A}^{\imath}(\theta)}\right\|_{\textstyle\mid\Sigma} = {\rm rank}_{
\varepsilon_{\bar{J}}} \left\|
\frac{\delta_l \phantom {xxxx}}{\delta{\cal A}^{\jmath}(\theta_1)}
\frac{\delta_l Z[{\cal A}]}{\delta{\cal
A}^{A}(\theta)}\right\|_{\textstyle\mid\Sigma} = n-m\,.
\end{eqnarray}
Remind [1], in the first place, that $\Sigma$ is considered as local
supersurface  and, in the second,  the  following representation is true for
DCLF in terms of
the superfields $\tilde{\cal A}^{\imath}(\theta)$ = ${\cal A}^{\imath}(
\theta)$ $-$ ${\cal A}^{\imath}_0(\theta)$:
$\tilde{\cal A}^{\imath}_0(\theta)=0$ $\in$ $\Sigma$
\begin{eqnarray}
{} & {} & {\Theta}_{\imath} \bigl(
{\cal A}(\theta), {\stackrel{\ \circ}{\cal A}}(\theta), \theta \bigr) =
{\Theta}_{\imath\; {\rm lin}} \bigl(\tilde{\cal A}(\theta),
{\stackrel{\ \circ}{\tilde{\cal A}}}(\theta),
\theta \bigr) + {\Theta}_{\imath\;{\rm nl}} \bigl( \tilde{\cal A}(\theta),
{\stackrel{\ \circ}{\tilde{\cal A}}} (\theta),
\theta \bigr)\,,\nonumber \\
{} & {} & \Bigl(\min{\rm deg}_{\tilde{\cal A}(\theta){\stackrel{\
\circ}{\tilde{\cal A}}}(\theta)}, {\rm deg}_{\tilde{{\cal
A}}(\theta){\stackrel{\ \circ}{\tilde{\cal
A}}}(\theta)}\Bigr){\Theta}_{\imath\;{\rm lin}}(\theta) = (1, 1),\ \min{\rm
deg}_{\tilde{{\cal A}}(\theta){\stackrel{\
\circ}{\tilde{\cal A}}}(\theta)} {\Theta}_{\imath\;{\rm nl}}(\theta) \geq 2
\,.
\end{eqnarray}
Whereas the assumption $2$ gives the possibility to present $\Theta_{\imath
}(\theta)$ in the form of two special subsystems in formal
ignoring of the requirements of locality and covariance with respect to
index $\imath$ relative to restriction of the superfield representation
$T$ onto subsupergroup $\bar{J}: T_{\vert\bar{J}}$.

Reduction of DCLF to equivalent system of the 1st order on
$\theta$ ODE in GNF immediately follows from the Theorem 1 application
 [1] in the form of

\noindent
\underline{\bf Theorem 2}

\noindent
A nondegenerate parametrization for ${\cal A}^{\imath}(\theta)$ exists
\begin{eqnarray}
{} & {} & \hspace{-4em}{{\cal A}}^{\imath}(\theta) = \bigl(
\delta^{\bar{\imath}}(\theta), \beta^{\underline{\imath}} (\theta),
\xi^{\alpha}(\theta)\bigr) \equiv \bigl(\varphi^{A}(\theta), \xi^{
\alpha}(\theta)\bigr),\ \imath = (\bar{\imath}, {\underline{\imath}}, \alpha)
\equiv (A, \alpha),\ \bar{\imath} = 1,\ldots, n - \underline{m}, \nonumber \\
{} & {} & \hspace{-4em} {\underline{\imath}} = n- \underline{m}+1, \ldots,
n-m,\;\underline{m}=(\underline{m}_+ , \underline{m}_-),\
 A=1,\ldots,n-m,\;\alpha = n-m+1,\ldots,n, 
\end{eqnarray}
the such that $\Theta_{\imath}(\theta)$ (3.3) are
equivalent to the system of independent ODE in GNF
\begin{eqnarray}
{\stackrel{\;\circ}{\delta}}{}^{\bar{\imath}}(\theta) =
\phi^{\bar{\imath}}
\bigl( \delta(\theta),{\stackrel{\;\circ}{\xi}}(\theta), \xi(\theta),
\theta\bigr),\ \;
{\beta}^{\underline{\imath}}(\theta) = \kappa^{\underline{\imath}}
 \bigl( \delta(\theta), \xi(\theta), \theta\bigr)\,,
\end{eqnarray}
with $\phi^{\bar{\imath}}(\theta), \kappa^{\underline{\imath}}(\theta)\in
C^k(T_{odd}{\cal M}_{cl}\times\{\theta\})$ and
arbitrary superfields $\xi^{\alpha}(\theta)$: $[\xi^{\alpha}(\theta)]$ = $m
<n$. Their $(\xi^{\alpha}(\theta))$ number coincides with one of differential
identities among $\Theta_{\imath}(\theta)$
\begin{eqnarray}
{}\hspace{-1em}\int
\hspace{-0.4em} d\theta {}\displaystyle\frac{\delta Z[{\cal A} ]}{\delta
{\cal A}^{\imath}(\theta)}{}{\hat{{\cal R}}}^{\imath}_{\alpha} \bigl(\theta;
{\theta}' \bigr)\hspace{-0.1em}= \hspace{-0.1em}
0,\ (\varepsilon_{P}, \varepsilon_{\bar{J}}, \varepsilon) {\hat{{\cal
R}}}^{\imath}_{\alpha}(\theta; {\theta}') \hspace{-0.1em}=\hspace{-0.1em} (1
\hspace{-0.1em}+\hspace{-0.1em}
(\varepsilon_P)_{\imath}, (\varepsilon_{\bar{J}})_{\imath}\hspace{-0.1em}+
\hspace{-0.1em}\varepsilon_{\alpha} , \varepsilon_{\imath} \hspace{-0.1em}+
\hspace{-0.1em}\varepsilon_{\alpha} \hspace{-0.1em}+\hspace{-0.1em}1)
\end{eqnarray}
with a) local and b)  functionally independent operators
${\hat{\cal R}}^{\imath}_{\alpha}
\bigl({\cal A}(\theta)$, ${\stackrel{\ \circ}{\cal A}}(\theta),
\theta;\theta'\bigr)$ $\equiv$ ${\hat{\cal R}}^{\imath}_{\alpha}(\theta;
{\theta}')$:
\begin{eqnarray}
{}\hspace{-6.5cm} {\rm a)}
\hspace{3.0cm}{\hat{\cal R}}^{\imath}_{\alpha}(\theta; {\theta}') =
\sum\limits_{k=0}^{1} \left(\left(\displaystyle\frac{d}{d\theta} \right)^k
\delta(\theta - \theta')\right)
{\hat{ {\cal R}}}_{k}{}^{\imath}_{\alpha}
\bigl( {\cal A}(\theta), {\stackrel{\ \circ}{\cal A}}(\theta), \theta\bigr),\
\\ 
\hspace{-3cm}
\phantom{\rm a)} \hspace{2.5cm}(\varepsilon_{P}, \varepsilon_{\bar{J}},
\varepsilon) {\hat{{\cal R}}}_{k}{}^{\imath}_{ \alpha}(\theta) =
 (\delta_{1k} +
(\varepsilon_P)_{\imath}, (\varepsilon_{\bar{J}})_{\imath}+
\varepsilon_{\alpha} ,
\varepsilon_{\imath}+ \varepsilon_{\alpha} + \delta_{1k}),\;k=0,1\;, 
\end{eqnarray}
b) functional equation
\begin{eqnarray}
\int d\theta' {\hat{{\cal R}}}^{\imath}_{\alpha}(\theta;
{\theta}') u^{\alpha}\bigl( {\cal A}(\theta'), {\stackrel{\
\circ}{\cal A}}(\theta'), \theta'\bigr) = 0,\ u^{\alpha}(\theta) \in
C^k(T_{odd}{\cal M}_{cl}\times\{\theta\}) 
\end{eqnarray}
has the unique vanishing solution.
\vspace{1ex}

It literally follows from Theorem 2, after change of corresponding
symbols, the consequence being analogous to Corollary 1 for HCLF
$\Theta_{\imath}({\cal A}(\theta),\theta)$   with
${\cal R}_{0}{}^{\imath}_{ \alpha}({\cal A}(\theta),\theta)$ = ${\hat{
\cal R}}_{0}{}^{\imath}_{\alpha}({\cal A}(\theta),
\theta)$ [1].

The interpretation for  operators
${\hat{\cal R}}^{\imath}_{\alpha}(\theta;{\theta}')$,
${\cal R}_{0}{}^{\imath}_{ \alpha}({\cal A}(\theta),\theta)$
as the GGTGT, GGTST respectively had been given in Ref.[1]. It had been
shown the complete sets of the GGTGT, GGTST appear by the bases in affine
$C^k(T_{odd}{\cal M}_{cl}\times\{\theta\})$-module $Q(Z)$ = ${\rm Ker}\{
\frac{\delta_l Z[{\cal A}]\phantom{x}}{\delta{\cal A}^{\imath}(\theta)}\}$ and
affine $C^k({\cal M}_{cl}\times\{\theta\})$-module
$Q(S_L)$ = ${\rm Ker}\{\Theta_{\imath}({\cal A}(\theta),\theta)\}$
respectively. By realization of the mentioned consequence for Theorem 2 it
appears the Corollary 2.2 from Ref.[1] in the
framework of which a GSTF model is the almost natural system
\begin{eqnarray}
{} & S_L \bigl( {\cal A}(\theta), {\stackrel{\
\circ}{\cal A}}(\theta), \theta\bigr) = T \bigl( {\cal A}(\theta),
{\stackrel{\ \circ}{\cal A}}(\theta)\bigr) - S\bigl( {\cal A}(\theta),
 \theta\bigr),\ \min{\rm deg}_{{\cal A}(\theta)}S(\theta) = 2 \;, & {}
\\ 
{} & T\bigl( {\cal A}(\theta), {\stackrel{\ \circ}{\cal A}}(\theta)\bigr)
= T_1\bigl({\stackrel{\ \circ}{\cal A}}(\theta)\bigr) +
{\stackrel{\;\circ}{{\cal A}^{\jmath}}}(\theta) T_{\jmath}\bigl( {\cal A}
(\theta)\bigr),\ T_{\jmath}\bigl( {\cal A}(\theta)\bigr) = g_{\jmath
\imath}(\theta){\cal A}^{\imath} (\theta)\ , & {} \nonumber\\
{} &  g_{\jmath
\imath}(\theta) = (-1)^{\varepsilon_{\jmath}\varepsilon_{\imath}}g_{\imath
\jmath}(\theta),\ g_{\jmath \imath}(\theta) = P_0(\theta)g_{\jmath
\imath}(\theta),\ \min{\rm deg}_{{\stackrel{\;\circ}{\cal
A}}(\theta)}T_1(\theta) = 2\,, & {} 
\end{eqnarray}
so that the HCLF and condition (3.7) have the form
\begin{eqnarray}
{\Theta}_{\imath}
\bigl({\cal A}(\theta), \theta \bigr) = - S,_{\imath} \bigl( {{\cal
A}}(\theta), \theta \bigr)(-1)^{ \varepsilon_{\imath}} = 0,\
{\rm rank}_{\varepsilon_{\bar{J}}}\left\|S,_{\imath
\jmath} \bigl( {{\cal A}}(\theta),\theta)\right\|_{\textstyle \mid \Sigma}
= n-m\,, 
\end{eqnarray}
Corresponding GTGT, GTST have the form  [1]
\begin{eqnarray}
{} &
{\cal A}^{\imath}(\theta) \mapsto {\cal A}'^{\imath
}(\theta) = {\cal A}^{\imath}(\theta) + \delta_g{\cal A}^{\imath}(\theta);\
\delta_g{\cal A}^{\imath}(\theta) = \displaystyle\int d\theta'
\hat{\cal R}^{\imath}_{\alpha}(\theta;\theta')\xi^{\alpha}(\theta')
\,,{} &
\\
{} & {\cal A}^{\imath}(\theta) \mapsto {\cal A}'^{\imath}(\theta) =
{\cal A}^{\imath}(\theta) + \delta{\cal A}^{\imath}(\theta);\
\delta{\cal A}^{\imath}(\theta) =
{\cal R}_0{}^{\imath}_{\alpha}({\cal A}(\theta),\theta)\xi_0^{\alpha}(
\theta){} &
\end{eqnarray}
and appear by infinitesimal invariance transformations with arbitrary
superfunctions $\xi^{\alpha}(\theta)$, $\xi_0^{\alpha}(\theta)$
[$(\varepsilon_{P}, \varepsilon_{\bar{J}}, \varepsilon)\xi^{\alpha}(\theta)$
= $(0, \varepsilon_{\alpha}, \varepsilon_{\alpha})$]
for $
Z[{\cal A}]$, $S({\cal A}(\theta), \theta)$ respectively.

In addition for local with respect to $z^a$ models the GGTGT, GGTST can be
represented by the local differential operators with respect to $z^a$.
At least in ignoring of the requirement of covariance on index $\imath$ the
GGTGT, GGTST, as it follows from Theorem 2, appear by independent and hence
define the irreducible GSTF models in the Lagrangian
formulation, called the irreducible GThGT and GThST (gauge theory of the
special type) respectively [1]. In general case the conservation of
mentioned conditions on locality and covariance for
${\hat{\cal R}}^{\imath}_{\alpha}(\theta; {\theta}')$,
${\cal R}_{0}{}^{\imath}_{ \alpha}(\theta)$
leads to modification of the Theorem 2 conclusion
concerning of the solution for Eq.(3.14) and its analog for
${\cal R}_{0}{}^{\imath}_{ \alpha}(\theta)$.
Namely, for the last equations the nonvanishing solutions may exist as well.
By definition
the GThGT (GThST) with the such property are called the reducible GThGT
(GThST) with functionally dependent GGTGT (GGTST).
\section{Gauge Algebra of GGTGT}
\setcounter{equation}{0}

Following to  Refs.[6,7] let us investigate the GTGT (3.18) and algebraic
structures connected with them. From GTGT in a more than one-valued form one
can construct the finite transformations of invariance for $Z[{\cal A}]$
\begin{eqnarray}
{} & {\cal A}^{\imath}(\theta) \mapsto {\cal A}^{\imath}_f(\theta)=
G^{\imath}({\cal A}(\theta)\vert \xi(\theta)),\
G^{\imath}({\cal A}(\theta)\vert 0)={\cal A}^{\imath}(\theta)\,, {} &
\\
{} & \hspace{-0.5em}\displaystyle\frac{\delta_l G^{\imath}({
\cal A}(\theta)\vert
\xi(\theta))}{\delta\xi^{\alpha}(\theta')\phantom{xxxxxx}}{\hspace{-0.5em}
\phantom{\Bigr)}}_{\mid\xi^{\alpha}(\theta)=0}=
\hat{\cal R}^{\imath}_{\alpha}(\theta;\theta'):
Z[{\cal A}_f] = Z[{\cal A}]
,\,(\varepsilon_{P}, \varepsilon_{\bar{J}}, \varepsilon)\displaystyle\frac{
\delta_l\phantom{xxx}}{\delta\xi^{\alpha}(\theta)}\hspace{-0.1em}=
\hspace{-0.1em}
(1, \varepsilon_{\alpha}, \varepsilon_{\alpha}+1)\,.{} & 
\end{eqnarray}
As $G^{\imath}(\theta)$ one can make use, for instance, the superfields
satisfying to the $\theta$-superfield condition [6]
\begin{eqnarray}
\frac{\partial }{\partial\tau}G^{\imath}\bigl({\cal A}(\theta)\vert\tau
\xi(\theta)\bigr)=
\int d\theta' \xi^{\alpha}(\theta'){\hat{{\cal R}}}^{\imath}_{\alpha}(\theta;
{\theta}')_{\mid
{\cal A}^{\imath}(\theta) =
G^{\imath}({\cal A}(\theta)\mid\tau\xi(\theta))},\ \tau \in
{\bf R}\,.
\end{eqnarray}
Really, having denoted $Z_{\tau}\equiv Z[{\cal A}]_{\mid{\cal A}=G({\cal A}
\vert\tau\xi)}$, obtain from (3.11), (4.3) the relationships
\begin{eqnarray}
\frac{\partial }{\partial\tau}Z_{\tau}=
\int d\theta  \frac{\delta Z[{\cal A} ]}{\delta
{\cal A}^{\imath}(\theta)}
\int d\theta' \xi^{\alpha}(\theta'){\hat{{\cal R}}}^{
\imath}_{\alpha}(\theta;{\theta}'){\hspace{-0.5em}\phantom{\Bigr)}}_{
\mid{\cal A}^{\imath}(\theta) =
G^{\imath}({\cal A}(\theta)\vert\tau\xi(\theta))}=0\,,
\end{eqnarray}
from which it follows
\begin{eqnarray}
Z_{\tau=0}=Z_{\tau=1} \Longleftrightarrow  Z[{\cal A}]=Z[{\cal A}_f]\,.
\end{eqnarray}
Formal solution for Eq.(4.3) with initial condition from (4.1) has the form
\begin{eqnarray}
{} & G^{\imath}\bigl({\cal A}(\theta)\vert\xi(\theta)\bigr)=
{\rm exp}\bigl\{
\displaystyle\int\hspace{-0.4em} d\theta'\xi^{\alpha}(\theta')\hat{\Gamma}_{
\alpha}(\theta')\bigr\}{\cal A}^{\imath}(\theta)\,,{} & \\
{} & \hat{\Gamma}_{\alpha}(\theta')F[{\cal A}]=
\displaystyle\int  d\theta
 \frac{\delta F[{\cal A} ]}{\delta
{\cal A}^{\imath}(\theta)}{\hat{{\cal R}}}^{\imath}_{\alpha}(\theta;\theta'),\
(\varepsilon_{P}, \varepsilon_{\bar{J}}, \varepsilon)
\hat{\Gamma}_{\alpha}(\theta)= (1, \varepsilon_{\alpha},
\varepsilon_{\alpha}+1),\; F[{\cal A} ] \in C_F \,.
\end{eqnarray}
Really, for an arbitrary superfunctional $F[{\cal A}]$, having the polynomial
series expansion with respect to ${\cal A}^{\imath}(\theta)$, the operatorial
formula is valid
\begin{eqnarray}
F[G]\equiv F[{\cal A}_f]= {\rm exp}\bigl\{
\int d\theta'\xi^{\alpha}(\theta')\hat{\Gamma}_{\alpha}(\theta')
\bigr\}F[{\cal A}]\,.
\end{eqnarray}
Choosing  $F[G]\equiv G^{\imath}({\cal A}(\theta)\vert\xi(\theta))$
obtain the solution of Eqs.(4.2) in the form (4.6).

From differential consequences of the identities (3.11)
\begin{eqnarray}
\left(\int d\theta\Bigl[\Bigl(
\frac{\delta\phantom{xxx}}{\delta{\cal A}^{\imath}(\theta)}
\frac{\delta Z[{\cal A}]\phantom{
x}}{\delta{\cal A}^{\jmath}(\theta'')}\Bigr)
{\hat{{\cal R}}}^{\imath}_{\alpha}(\theta;\theta')(-1)^{\varepsilon_{\jmath}
\varepsilon_{\alpha}} - \frac{\delta Z[{\cal A}]\phantom{
x}}{\delta{\cal A}^{\imath}(\theta)}
\frac{\delta{\hat{{\cal R}}}^{\imath}_{\alpha}(\theta;\theta') }{
\delta{\cal A}^{\jmath}(\theta'')\phantom{xx}}\Bigr]\right)(-1)^{
\varepsilon_{\imath}}=0
\end{eqnarray}
it follows the transformation rule for $\frac{\delta Z[{\cal A}]\phantom{}
}{\delta{\cal A}^{\imath}(\theta)}$ under finite GTGT (4.1), (4.6)
\begin{eqnarray}
\frac{\delta Z[{\cal A}]\phantom{}}{\delta{\cal A}^{\imath}(\theta)}{
\hspace{-0.5em}\phantom{\Bigr)}}_{\mid {\cal A}^{\imath} =
G^{\imath}({\cal A}\vert\xi)} =
\int d\theta' Q_{\imath}{
}^{\jmath}\bigl({\cal A}(\theta), {\stackrel{\ \circ}{\cal A}}(\theta),
\theta; \theta'\bigr)
\frac{\delta Z[{\cal A}]\phantom{}}{\delta{\cal A}^{\jmath}(\theta')}
\end{eqnarray}
with nondegenerate supermatrix $Q_{\imath}{}^{\jmath}(\theta;\theta')$ $\in$
$C^k(T_{odd}{\cal M}_{cl}\times\{\theta,\theta'\})$
in a some neighbourhood of $\xi^{\alpha}(\theta)$ = $0$: $
Q_{\imath}{}^{\jmath}(\theta;\theta')_{\mid\xi(\theta)=0}$ = $\delta(\theta'-
\theta)\delta_{\imath}{}^{\jmath}$.

Investigation  of the gauge algebra of  GTGT properties is based on the
study of properties of the supercommutator  of the 1st order
differential operators $\hat{\Gamma}_{\alpha}(\theta)$, with respect to
variational superfield derivatives on ${\cal A}^{\imath}(\theta)$ (4.7),
having
$Z[{\cal A}]$ as the eigensuperfunction with zero eigenvalue. By definition,
the supercommutator $[\hat{\Gamma}_{\alpha}(\theta_1), \hat{\Gamma}_{
\beta}(\theta_2)]_s$ possesses by the same property as well.
Its value in calculating on arbitrary $F[{\cal A}]$ $\in$ $C_F$ is equal to
\begin{eqnarray}
{} &
[\hat{\Gamma}_{\alpha}(\theta_1), \hat{\Gamma}_{\beta}(\theta_2)]_s
F[{\cal A}]\hspace{-0.1em}=\hspace{-0.1em}
\hat{\Gamma}_{\alpha}(\theta_1)\left(\hat{\Gamma}_{\beta}(\theta_2)F[{\cal A}]
\right)\hspace{-0.1em} -\hspace{-0.1em}
(-1)^{(\varepsilon_{\alpha}+1)(\varepsilon_{\beta}+1)}\bigl((\alpha,\theta_1)
\hspace{-0.1em}
\longleftrightarrow\hspace{-0.1em} (\beta,\theta_2)\bigr)\hspace{-0.1em} =
\hspace{-0.1em}{} &\nonumber \\
{} &
\displaystyle\int\hspace{-0.2em} d\theta_2'
\displaystyle\frac{\delta F[{\cal A}]\phantom{
x}}{\delta{\cal A}^{\jmath}(\theta_2')}\displaystyle\int\hspace{-0.2em}
d\theta_1'\left(\Bigl(\frac{\delta{\hat{\cal R}}^{\jmath}_{\beta}(\theta_2';
\theta_2)}{\delta{\cal A}^{\imath}(\theta_1')\phantom{xx}}\Bigr)
{\hat{\cal R}}^{\imath}_{\alpha}(\theta_1';\theta_1)\hspace{-0.2em} -
\hspace{-0.2em}
(-1)^{(\varepsilon_{\alpha}+1)(\varepsilon_{\beta}+1)}\bigl((\alpha,\theta_1)
\hspace{-0.2em}\longleftrightarrow \hspace{-0.2em}
(\beta,\theta_2)\bigr) \hspace{-0.2em}\right) {} &
\nonumber \\
{} &
=\hspace{-0.2em}
\displaystyle\int d\theta_2'\displaystyle\frac{\delta F[{\cal A}]\phantom{
x}}{\delta{\cal A}^{\jmath}(\theta_2')}
\hat{y}^{\jmath}_{\beta\alpha}\bigl({\cal A}(\theta_2'), {\stackrel{\
\circ}{\cal A}}(\theta_2'), \theta_2'; \theta_2, \theta_1\bigr)\,,{} &
\\
{} & \hspace{-0.7em}
\hat{y}^{\imath}_{\beta\alpha}(\theta_1'; \theta_2, \theta_1)\hspace{-0.2em}=
\hspace{-0.2em}-\hspace{-0.1em} (-1)^{(\varepsilon_{\alpha}+1)(\varepsilon_{
\beta}+1)}\hat{y}^{\imath}_{\alpha\beta}(\theta_1'; \vec{\theta}_2),\,
(\varepsilon_{P}, \hspace{-0.1em}\varepsilon_{\bar{J}}, \hspace{-0.1em}
\varepsilon)\hat{y}^{\imath}_{
\beta\alpha}\hspace{-0.15em}=\hspace{-0.15em}
((\varepsilon_P)_{\imath}, (\varepsilon_{\bar{J}})_{\imath}\hspace{-0.15em} +
\hspace{-0.15em}\varepsilon_{\alpha}
\hspace{-0.15em}+\hspace{-0.15em} \varepsilon_{\beta},
\varepsilon_{\imath}\hspace{-0.15em} +
\hspace{-0.15em}\varepsilon_{\alpha} \hspace{-0.15em}+ \hspace{-0.15em}
\varepsilon_{\beta}),{} &\nonumber \\
{} & \hat{y}^{\imath}_{\alpha\beta}\bigl({\cal A}(\theta_1'), {\stackrel{\
\circ}{\cal A}}(\theta_1'), \theta_1'; \theta_1,\theta_2\bigr) \equiv
\hat{y}^{\imath}_{\alpha\beta}(\theta_1'; \vec{\theta}_2) \in
C^k(T_{odd}{\cal M}_{cl}\times\{\theta_1',\vec{\theta}_2\}),
\vec{\theta}_k \equiv \theta_1,\ldots,\theta_k\,.{} &
\end{eqnarray}
Superfunctions $\hat{y}^{\imath}_{\alpha\beta}(\theta'_1;\vec{\theta}_2)$
appear by local operators of differentiation on $\theta$ (and with respect to
$z^a$, if ${\hat{\cal R}}^{\imath}_{\alpha}(\theta; {\theta}')$ are the same).
By virtue of completeness of the GGTGT $
{\hat{\cal R}}^{\imath}_{\alpha}(\theta; {\theta}')$ the quantities
$\hat{y}^{\imath}_{\alpha\beta}$ must be expressed through GGTGT and the
trivial GGTGT  $\hat{\tau}^{\imath}_{\alpha\beta}(\theta'_1;\vec{\theta
}_2)$ [1]
\begin{eqnarray}
\hat{y}^{\imath}_{\alpha\beta}(\theta_1';\vec{\theta}_2) =
(-1)^{\varepsilon_{\imath}}\int d\theta_3
{\hat{\cal R}}^{\imath}_{\gamma}(\theta_1';\theta_3)
\hat{\cal F}^{\gamma}_{\alpha\beta}(\theta_3;\vec{\theta}_2) +
\int d\theta_2'\frac{\delta Z[{\cal A}]\phantom{
x}}{\delta{\cal A}^{\jmath}(\theta_2')}
\hat{\cal M}^{\imath\jmath}_{\alpha\beta}(\vec{\theta}{}_2';\vec{\theta}_2)
(-1)^{\varepsilon_{\jmath}}\,.{} & 
\end{eqnarray}
Superfunctions  $\hat{\cal F}^{\gamma}_{\alpha\beta}(\theta_3;\vec{\theta
}_2)$, $\hat{\cal M}^{\imath\jmath}_{\alpha\beta}(\vec{\theta}{}'_2;
\vec{\theta}_2)$ $\in$ $C^k\bigl(T_{odd}{\cal M}_{cl}\times\{\vec{\theta}_3,
\vec{\theta}{}'_2\}\bigr)$ possess by the  properties
\begin{eqnarray}
{} & \begin{array}{l|ccc}
{}& \varepsilon_P & \varepsilon_{\bar{J}} & \varepsilon \\ \hline
\hat{\cal F}^{\gamma}_{\alpha\beta}(\theta_3;\vec{\theta}_2) & 0    &
\varepsilon_{\gamma} + \varepsilon_{\alpha}+ \varepsilon_{\beta} &
\varepsilon_{\gamma} + \varepsilon_{\alpha}+ \varepsilon_{\beta} \\
\hat{\cal M}^{\imath\jmath}_{\alpha\beta}(\vec{\theta}{}'_2;\vec{\theta}_2) &
1+(\varepsilon_P)_{\imath}+(\varepsilon_P)_{\jmath} &
(\varepsilon_{\bar{J}})_{\imath} + (\varepsilon_{\bar{J}})_{\jmath} +
\varepsilon_{\alpha}+
\varepsilon_{\beta} & 1+
\varepsilon_{\imath} + \varepsilon_{\jmath} + \varepsilon_{\alpha}+
\varepsilon_{\beta}
\end{array}, {} & \\
{} & \hat{\cal F}^{\gamma}_{\alpha\beta}(\theta_3;\vec{\theta}_2) \equiv
\hat{\cal F}^{\gamma}_{\alpha\beta}\bigl({\cal A}(\theta_3), {\stackrel{\
\circ}{\cal A}}(\theta_3), \theta_3; \vec{\theta}_2\bigr) =
- (-1)^{(\varepsilon_{\alpha}+1)(\varepsilon_{\beta}+1)}
\hat{\cal F}^{\gamma}_{\beta\alpha}(\theta_3;\theta_2,\theta_1)\,,{} &
\nonumber \\
{} &
\hat{\cal M}^{\imath\jmath}_{\alpha\beta}(\vec{\theta}{}'_2;\vec{\theta}_2)
\equiv
\hat{\cal M}^{\imath\jmath}_{\alpha\beta}
\bigl({\cal A}(\theta_1'), {\stackrel{\
\circ}{\cal A}}(\theta_1'), \vec{\theta}{}'_2;\vec{\theta}_2\bigr) =
- (-1)^{(\varepsilon_{\imath}+1)(\varepsilon_{\jmath}+1)}
\hat{\cal M}^{\jmath\imath}_{\alpha\beta}(\theta'_2,\theta'_1;
\vec{\theta}_2) {} & \nonumber\\
{} &
= - (-1)^{(\varepsilon_{\alpha}+1)(\varepsilon_{\beta}+1)}
\hat{\cal M}^{\imath\jmath}_{\beta\alpha}(\vec{\theta}{}'_2;{\theta}_2,
{\theta}_1) {} & 
\end{eqnarray}
and can be chosen by $\theta$-local ones.

The explicit form of $\hat{\cal F}^{\gamma}_{\alpha\beta}$,
$\hat{\cal M}^{\imath\jmath}_{\alpha\beta}$ and their properties are based on
the analysis of the general solution for equation
\begin{eqnarray}
\int d\theta
\frac{\delta Z[{\cal A}]\phantom{x}}{\delta{\cal A}^{\imath}(\theta)}
\hat{y}^{\imath}\bigl( {\cal A}(\theta), {\stackrel{\ \circ}{\cal A}
}(\theta),\theta\bigr)=0
\,.
\end{eqnarray}
\underline{\bf Lemma 1}:

\noindent
General solution of the Eq.(4.16) for irreducible
GGTGT satisfying to completeness condition has the
form in the superalgebra $C^k\bigl(T_{odd}{\cal M}_{cl}\times\{\theta\})$
\begin{eqnarray}
{} &
\hat{y}^{\imath}(\theta) \equiv
\hat{y}^{\imath}\bigl( {\cal A}(\theta), {\stackrel{\ \circ}{\cal
A}}(\theta),\theta\bigr) = \displaystyle\int d\theta'\Bigl[
{\hat{\cal R}}^{\imath}_{\alpha}(\theta;\theta'){\hat{\Phi}}^{\alpha}(\theta'
)(-1)^{\varepsilon_{\imath}} +
\displaystyle\frac{\delta Z[{\cal A}]\phantom{x}}{\delta{\cal
A}^{\jmath}(\theta')}\hat{E}^{\imath\jmath}(\theta,\theta')(-1)^{\varepsilon_{
\jmath}}\Bigr]\,,{} &
\\
{} & \hspace{-1.0em}{\hat{\Phi}}^{\alpha}(\theta) \hspace{-0.1em}\equiv
\hspace{-0.1em}{\hat{\Phi}}^{\alpha}
\bigl( {\cal A}(\theta), {\stackrel{\ \circ}{\cal A}
}(\theta),\theta\bigr),\;
\hat{E}^{\imath\jmath}(\theta,\theta') \equiv
\hat{E}{}^{\imath\jmath}\bigl({\cal A}(\theta), {\stackrel{\ \circ}{\cal
A}}(\theta),\theta,\theta'\bigr) = -(-1)^{(\varepsilon_{\imath} +1)(
\varepsilon_{\jmath} +1)}\hat{E}^{\jmath\imath}(\theta',\theta)\,, {} &
\nonumber \\
{} &
\begin{array}{l|cccc}
{} & \varepsilon_P & \varepsilon_{\bar{J}}& \varepsilon & {}\\ \hline
{\hat{\Phi}}^{\alpha}(\theta) &
\varepsilon_{P}(\hat{y}^{\imath})+  (\varepsilon_{P})_{\imath} &
\varepsilon_{\bar{J}}(\hat{y}^{\imath}) +
(\varepsilon_{\bar{J}})_{\imath} + \varepsilon_{\alpha} &
\varepsilon(\hat{y}^{\imath}) + \varepsilon_{\imath}  +
\varepsilon_{\alpha} & {} \\
{\hat{E}}^{\imath\jmath}(\theta,\theta')  & \varepsilon_{P}(\hat{y}^{
\imath})+ (\varepsilon_{P})_{\jmath}+ 1 &
\varepsilon_{\bar{J}}(\hat{y}^{\imath}) + (\varepsilon_{\bar{J}})_{\jmath} &
\varepsilon(\hat{y}^{\imath}) + \varepsilon_{\jmath} + 1 & .\\
\end{array} 
\end{eqnarray}
\underline{Proof:} Assumption (3.7) permits to represent Eq.(4.16) and
identities (3.11) in the form respectively
\begin{eqnarray}
{} & {} &
\displaystyle\int d\theta\Bigl[
\frac{\delta Z[{\cal A}]\phantom{x}}{\delta{\cal A}^{A}(\theta)}
\hat{y}^{A}(\theta) +
\displaystyle\frac{\delta Z[{\cal A}]\phantom{x}}{\delta{\cal
A}^{\alpha}(\theta)}\hat{y}^{\alpha}(\theta)\Bigr]=0\,,
\\
{} & {} &
\displaystyle\int d\theta\Bigl[
\frac{\delta Z[{\cal A}]\phantom{x}}{\delta{\cal A}^{A}(\theta)}
\hat{\cal R}^{A}_{\beta}(\theta;\theta') + \displaystyle\frac{\delta
Z[{\cal A}]\phantom{x}}{\delta{\cal A}^{\alpha}(\theta)}
\hat{\cal R}^{\alpha}_{\beta}(\theta;\theta')\Bigr]=0
\,,\\
{} & {} &
\hat{\cal R}^{\imath}_{\beta}(\theta;\theta') \equiv \left(\hat{\cal
R}^{A}_{\beta}(\theta;\theta'), \hat{\cal R}^{\alpha}_{\beta}(\theta;\theta')
\right),\
{\rm rank}_{{\varepsilon}_{\bar{J}}}\left\|\hat{\cal R}^{\alpha}_{\beta}(
\theta;\theta')\right\|_{\mid\Sigma}\equiv \nonumber \\
{} & {} &
{\rm rank}_{{\varepsilon}_{\bar{J}}}\left\|\sum\limits_{k\geq 0}{\hat{
\cal R}_k}{}^{\alpha}_{\beta}(
\theta)\left(\frac{d}{d\theta}\right)^k\right\|_{\mid\Sigma}\delta(\theta-
\theta'):
{\rm rank}_{{\varepsilon}_{\bar{J}}}\left\|\sum\limits_{k\geq 0}{
\hat{\cal R}_k}{}^{
\alpha}_{\beta}(\theta)\left(\frac{d}{d\theta}\right)^k\right\|_{\mid\Sigma}=m
\,,\\
{} & {} & \exists \bigl(\hat{\cal R}^{-1}\bigr)^{\beta}{}_{\gamma}(\theta;
\theta_1): \,\displaystyle\int d\theta'
\hat{\cal R}^{\alpha}_{\beta}(\theta;\theta')
\bigl(\hat{\cal R}^{-1}\bigr)^{\beta}{}_{\gamma}(\theta';\theta_1)=\delta^{
\alpha}{}_{\gamma}\delta(\theta-\theta_1)\,.
\end{eqnarray}
From (4.19)--(4.22) it follows the equivalent representation for (3.11),
(4.16)
\begin{eqnarray}
{} & \displaystyle\frac{\delta Z[{\cal A}]\phantom{x}}{\delta{\cal
A}^{\gamma}(\theta_1)} = \displaystyle\int d\theta' d\theta
\frac{\delta Z[{\cal A}]\phantom{x}}{\delta{\cal A}^{B}(\theta)}
\hat{\cal R}^{B}_{\alpha}(\theta;\theta')
\bigl(\hat{\cal R}^{-1}\bigr)^{\alpha}{}_{\gamma}(\theta';\theta_1)
\,,{} & \\
{} &  \displaystyle\int d\theta d\theta_1
\frac{\delta Z[{\cal A}]\phantom{x}}{\delta{\cal
A}^{A}(\theta_1)} \hat{z}^A(\theta,\theta_1)(-1)^{\varepsilon_{A}}=0\,,{} &
\nonumber \\
{} & \hat{z}^A(\theta,\theta_1) = \delta(\theta_1-\theta)\hat{y}^A(\theta_1)
 - \displaystyle\int d\theta'\hat{\cal R}^{A}_{\alpha}(\theta_1;\theta')
\bigl(\hat{\cal R}^{-1}\bigr)^{\alpha}{}_{\gamma}(\theta';\theta)
\hat{y}^{\gamma}(\theta)\,.{} & 
\end{eqnarray}
Condition (3.7) guarantees the existence of the special parametrization for
${\cal A}^{\imath}(\theta)$
\begin{eqnarray}
{\cal A}^{\imath}(\theta)\mapsto \tilde{\cal A}{}^{\imath}(\theta)=
\left(\frac{\delta Z[{\cal A}]\phantom{x}}{\delta{\cal A}^{A}(\theta)},
{\cal A}^{\alpha}(\theta)\right)\equiv ({\cal F}_A(\theta),
{\cal A}^{\alpha}(\theta))\,,
\end{eqnarray}
in terms of which Eq.(4.24) is written in the form
\begin{eqnarray}
\int d\theta d\theta_1{\cal F}_A(\theta_1)
\hat{z}^A\bigl(\tilde{\cal A}(\theta),{\stackrel{\ \circ}{\tilde{\cal A}}}(
\theta), \theta,\theta_1\bigr)(-1)^{\varepsilon_{A}}=0\,.
\end{eqnarray}
Calculating the variational superfield derivative of expression (4.26) with
respect to  ${\cal F}_B(\theta_2)$ we obtain
\begin{eqnarray}
{} & \hat{z}^B(\theta,\theta_2) + \displaystyle\int d\theta_1
{\cal F}_A(\theta_1)
\displaystyle\frac{\delta_l \hat{z}^B(\theta,\theta_2)}{\delta{\cal
F}_{A}(\theta_1)\phantom{x}} =
\displaystyle\int d\theta_1{\cal F}_A(\theta_1){\cal P}^{BA}(
\theta;{\theta}_2,{\theta}_1)\,,\nonumber \\
{} & \displaystyle\frac{\delta_l {\cal F}_B(\theta_1)}{\delta{\cal
F}_{C}(\theta)\phantom{x}} = \delta_B{}^C\delta(\theta-\theta_1)(-1)^{
\varepsilon_B}\,,{} &\\
{} & {\cal P}^{BA}(\theta;{\theta}_2,{\theta}_1)\equiv
{\cal P}^{BA}\bigl(\tilde{\cal A}(\theta),{\stackrel{\ \circ}{\tilde{\cal A}}
}(\theta), \theta;{\theta}_2,{\theta}_1)=
\displaystyle\frac{\delta_l \hat{z}^B(\theta,\theta_2)}{\delta{\cal
F}_{A}(\theta_1)\phantom{x}} - (-1)^{(\varepsilon_A +1)(\varepsilon_B +1)}
\times{} & \nonumber \\
{} & \left((A,\theta_1)\leftrightarrow (B,\theta_2)\right),\
{\cal P}^{AB}(\theta;\vec{\theta}_2) =
- (-1)^{(\varepsilon_A +1)(\varepsilon_B +1)}{\cal P}^{BA}(
\theta;{\theta}_2,{\theta}_1)\,.
\end{eqnarray}
After scaled transformation ${\cal F}_A(\theta)\mapsto\tau{\cal F}_A(\theta)$,
$\tau\in [0,1]$ $\subset {\bf R}$ in (4.27) this equation will  pass
into system of the 1st order on $\tau$ ODE
\begin{eqnarray}
{} & \displaystyle\frac{d}{d\tau}\left(
\tau\hat{z}^B\bigl(
\tau{\cal F}(\theta), {\cal A}^{\alpha}(\theta),
\tau{\stackrel{\,\circ}{\cal F}}(\theta),{\stackrel{\ \circ}{\cal A}
}{}^{\alpha}(\theta), \theta,\theta_1\bigr)\right)=
\displaystyle\int d\theta_2 \tau{\cal F}_A(\theta_2) \times \nonumber \\
{} &{\cal P}^{BA}\bigl(
\tau{\cal F}(\theta), {\cal A}^{\alpha}(\theta),
\tau{\stackrel{\,\circ}{\cal F}}(\theta),{\stackrel{\ \circ}{\cal A}
}{}^{\alpha}(\theta),\theta;\vec{\theta}_2\bigr)\,.{} & 
\end{eqnarray}
By direct integration of Eq.(4.29) with respect to $\tau$ along the segment
$[0,1]$ we obtain (the integral is regarded as improper one)
\begin{eqnarray}
{} & \hat{z}^B\bigl(\tilde{\cal A}(\theta),{\stackrel{\ \circ}{\tilde{\cal
A}}}(\theta), \theta,\theta_1\bigr) -  \lim\limits_{\tau\to 0}\tau
\hat{z}^B\bigl(
\tau{\cal F}(\theta), {\cal A}^{\alpha}(\theta),
\tau{\stackrel{\circ}{\cal F}}(\theta),{\stackrel{\ \circ}{\cal A}
}{}^{\alpha}(\theta),\theta,\theta_1\bigr) = {} & \nonumber \\
{} &
\displaystyle\int d\theta_2{\cal F}_A(\theta_2)
\displaystyle\int\limits_0^1 d\tau \tau
{\cal P}^{BA}\bigl(
\tau{\cal F}(\theta), {\cal A}^{\alpha}(\theta),
\tau{\stackrel{\,\circ}{\cal F}}(\theta),{\stackrel{\ \circ}{\cal A}
}{}^{\alpha}(\theta),\theta;\vec{\theta}_2\bigr)\,.{} & 
\end{eqnarray}
The boundedness of the solution for Eq.(4.16) near ${\cal F}_A(\theta)$ = $0$
and existence of the integral from right-hand side by hypothesis of the Lemma
mean the limit in the left of (4.30) is equal to $0$
and the general solution for (4.16) taking account of (4.24) has the form
\begin{eqnarray}
{} & \hat{y}^A(\theta) = \displaystyle\int d\theta_1
\left[\displaystyle\int d\theta'\hat{\cal R}^{A}_{\alpha}(\theta;\theta')
\bigl(\hat{\cal R}^{-1}\bigr)^{\alpha}{}_{\gamma}(\theta';\theta_1)
\hat{y}^{\gamma}(\theta_1)-\hat{z}^A(\theta_1,\theta)\right]= {} &
\nonumber\\
{} & \hspace{-1em}\displaystyle\int d\theta_1\left[
{\hat{\cal R}}^{A}_{\alpha}(\theta;\theta_1)
{\hat{\Phi}}^{\alpha}\bigl({\cal A}(\theta_1),{\stackrel{\ \circ}{{\cal
A}}}(\theta_1),\theta_1\bigr)
(-1)^{\varepsilon_{A}} +
\displaystyle\frac{\delta Z[{\cal A}]\phantom{x}}{\delta{\cal
A}^{B}(\theta_1)}
\hat{E}^{AB}\bigl({\cal A}(\theta),{\stackrel{\ \circ}{{\cal
A}}}(\theta),\theta,\theta_1\bigr)(-1)^{\varepsilon_{B}}\right],{} &
\\
{} & \hat{E}^{AB}\bigl({\cal A}(\theta),{\stackrel{\ \circ}{{\cal
A}}}(\theta),\theta,\theta_1\bigr)=
\displaystyle\int d\theta_2\int\limits_0^1 d\tau \tau {\cal P}^{AB}\bigl(
\tau{\cal F}(\theta_2), {\cal A}^{\alpha}(\theta_2),
\tau{\stackrel{\,\circ}{\cal F}}(\theta_2),{\stackrel{\ \circ}{\cal A}
}{}^{\alpha}(\theta_2),\theta_2;{\theta},{\theta}_1\bigr)\,,{} &
\\
{} &
{\hat{\Phi}}^{\alpha}\bigl({\cal A}(\theta),{\stackrel{\ \circ}{{\cal
A}}}(\theta),\theta\bigr) = (-1)^{\varepsilon_{\alpha}}
\displaystyle\int d\theta'
\bigl(\hat{\cal R}^{-1}\bigr)^{\alpha}{}_{\gamma}(\theta;\theta')
\hat{y}^{\gamma}(\theta') {} &
\end{eqnarray}
with arbitrary superfunctions $\hat{y}^{\gamma}(\theta)$.

Setting
\begin{equation}
\hat{E}^{\alpha B}(\vec{\theta}_2) = \hat{E}^{A\beta}(\vec{\theta}_2) =
\hat{E}^{\alpha\beta}(\vec{\theta}_2) = 0 
\end{equation}
we arrive to validity of the formula (4.17) with the properties
(4.18).

\noindent
\underline{\bf Definitions}:

{\bf 1)} Let us call the identities (3.11), expressions for supercommutator
GGTGT (4.13) the \underline{structural} \underline{equations} of the 1st
and 2nd orders
respectively of the \underline{gauge algebra of GTGT}. Call the
superfunctional $Z[{\cal A}]$; superfunctions ${\hat{\cal R}}^{\imath}_{
\alpha}(\theta; {\theta}')$;
$\hat{\cal F}^{\gamma}_{\alpha\beta}(\theta$; $\vec{\theta}_2)$, $
\hat{\cal M}^{\imath\jmath}_{\alpha\beta}(\vec{\theta}{}'_2$ ;$\vec{\theta
}_2)$ by the \underline{structural} \underline{superfunctions} of
 zero; 1st; 2nd orders respectively. The set of
quantities $Z[{\cal A}]$, ${\hat{\cal R}}^{\imath}_{\alpha}$,
$\hat{\cal F}^{\gamma}_{\alpha\beta}$, $\hat{\cal M}^{\imath\jmath}_{
\alpha\beta}$ and so on together
with structural equations let us call the gauge algebra of  GTGT
on $Q(Z)$. Thus, the order
of the structural superfunction and equation is equal to the number of free
lower indices in the nonzero function and equation respectively.

{\bf 2)} The rank $R$ of the gauge algebra of GTGT, by definition, is given
by the maximal number
of  free upper indices for the such structural superfunction from the set of
all structural superfunctions that corresponding numbers for other elements of
this set not greater than given one $(R\in {\bf Z}$, $R \leq \infty)$.
For $R=0$ the
GSTF model appears by nondegenerate theory of general type (ThGT) [1].

Further structural equations and superfunctions of the gauge algebra are
deduced
from systematic use of the definitions of GThGT, Lemma 1 and all preceding
structural equations and functions including their differential consequences
in analyzing of supercommutators of the form $[\hat{\Gamma}_{\alpha_1}(
\theta_1)$, $[\hat{\Gamma}_{\alpha_2}(\theta_2)$, $[\ldots[\hat{\Gamma}_{
\alpha_{k-1}}(\theta_{k-1})$, $\hat{\Gamma}_{\alpha_{k}}(\theta_{k})]
\ldots]]]$, $k\geq 3$. This investigation remains out the paper's scope.
Let us only point out the maximal numbers of different with respect to set
of upper indices on the structural functions and equations in the fixed $k$th
order of the gauge algebra of GTGT are equal to $[\frac{k}{2}]+1$ and $
[\frac{k+1}{2}]$ respectively.

Note the non-invariance of the definition for structural superfunctions and
rank of
gauge algebra because of  GGTGT are defined by ambiguously [1] with accuracy
up to equivalence transformations and in view of the fact that the form of
structural functions and equations depends on a choice of parametrization
for superfields ${\cal A}^{\imath}(\theta)$.
\section{Connection with Gauge Algebra of Irreducible \mbox{GThST}}
\setcounter{equation}{0}

\sloppy
\begin{sloppypar}
It had been shown in [1] the $Q(S_L)$ appears by the
$C^k({\cal M}_{cl}\times\{\theta,
\theta'\})$-submodule of the affine $C^k(T_{odd}{\cal M}_{cl}\times
\{\theta,\theta'\})$-module $Q(Z)$. By analogy with Sec.4 one can deduce the
basic relationships by means of the literal change of corresponding symbols
and operations and find the quantities defining a gauge algebra for
irreducible GThST on $Q(S_L)$ $\equiv$ $Q(S)$ $\equiv$ ${\rm Ker}\{S,_{
\imath}(\theta)\}$ with $S(\theta)$ (3.15) satisfying
to (3.17). Namely, the identities (3.11), the finite invariance
transformations  for $S({\cal A}(\theta), \theta)$ constructed from
infinitesimal
GTST (3.19)  analogously to scheme of Sec.4
(relationships (4.1)--(4.8)) and transformation rule for HCLF
$\Theta_{\imath}({\cal A}(\theta),\theta)$ under finite GTST have the form
respectively
\end{sloppypar}
\begin{eqnarray}
{} & {} & \hspace{-2.5em}
{S,}_{\imath}(\theta){{\cal R}_0}^{\imath}_{\alpha}(
{\cal A}(\theta),\theta)=0\,,\\
{} & {} & \hspace{-2.5em}
{\cal A}^{\imath}(\theta) \mapsto {\cal A}^{\imath}_{fin}(\theta)=
G^{\imath}_0({\cal A}(\theta)\vert \xi(\theta)),\
G^{\imath}_0({\cal A}(\theta)\vert 0)={\cal A}^{\imath}(\theta)\,,\\
{} & {} & \hspace{-2.5em}
\displaystyle\frac{\partial_l G^{\imath}_0({\cal A}(\theta)\vert
\xi(\theta))}{\partial\xi^{\alpha}(\theta)\phantom{xxxxxxx}}{\hspace{-0.5em}
\phantom{\Bigr)}}_{\mid\xi^{\alpha}(\theta)=0}=
{{\cal R}_0}^{\imath}_{\alpha}({\cal A}(\theta),\theta)
\,,\\
{} & {} & \hspace{-2.5em}
S({\cal A}_{fin}(\theta), \theta) = S({\cal A}(\theta),\theta)
\,,\\
{} & {} & \hspace{-2.5em}
G^{\imath}_0({\cal A}(\theta)\vert \xi(\theta)) = {\rm exp}\bigl\{
\xi^{\alpha}(\theta)\Gamma_{0{}\alpha}(\theta)\bigr\}{\cal A}^{\imath}(
\theta)\,,\\
{} & {} &\hspace{-2.5em}
\Gamma_{0{}\alpha}(\theta){\cal F}({\cal A}(\theta),\theta)=
{{\cal F},}_{\imath}({\cal A}(\theta),\theta)
{\cal R}_0{}^{\imath}_{\alpha}({\cal A}(\theta),\theta),\
(\varepsilon_P,\varepsilon_{\bar{J}},\varepsilon)\Gamma_{0{}\alpha}(\theta) =
(0,\varepsilon_{\alpha},\varepsilon_{\alpha})\,,\\
{} & {} &\hspace{-2.5em}
{\cal F}_{\imath}(G_0(\theta),\theta) \equiv {\cal F}_{\imath}({\cal
A}_{fin}(\theta),\theta) = {\rm exp}\bigl\{\xi^{\alpha}(\theta)
\Gamma_{0{}\alpha}(\theta)\bigr\}{\cal F}({\cal A}(\theta),\theta),\;
{\cal F}(\theta) \in C^k({\cal M}_{cl}\times\{\theta\})\,,\\
{} & {} &\hspace{-2.5em}
{S,}_{\imath}(\theta){\hspace{-0.5em}\phantom{\Bigr)}}_{\mid
{\cal A}^{\imath}(\theta) =
G^{\imath}_0({\cal A}(\theta)\vert \xi(\theta))}\hspace{-0.1em}
=\hspace{-0.1em} Q_0{}_{\imath}{}^{\jmath}({\cal A}(\theta),\theta)
{S,}_{\jmath}(\theta),\;{\rm sdet}\left\|Q_0{}_{\imath}{}^{\jmath}(\theta)
\right\|\hspace{-0.1em}\ne\hspace{-0.1em} 0,
Q_0{}_{\imath}{}^{\jmath}(\theta){\hspace{-0.5em}\phantom{\Bigr)}}_{
\mid\xi(\theta)=0}\hspace{-0.1em}=
\hspace{-0.1em}\delta_{\imath}{}^{\jmath}.
\end{eqnarray}
Vector  fields $\Gamma_{0{}\alpha}(\theta)$ on ${\cal M}_{cl}\times
\{\theta\}$, annulling $S({\cal A}(\theta),\theta)$, play  the role of
operators $\hat{\Gamma}_{\alpha}(\theta)$ in \mbox{GThGT} on
$T_{odd}{\cal M}_{cl}
\times \{\theta\}$. Their supercommutator possesses by the same property that
according to (4.11)--(4.15) means respectively
\begin{eqnarray}
{} & [\Gamma_{0{}\alpha}(\theta),\Gamma_{0{}\beta}(\theta)]_s
{\cal F}(\theta) = {{\cal F},}_{\jmath}(\theta)\Bigl(
{\cal R}_{0}{}^{\jmath}_{\beta},_{\imath}(\theta)
{\cal R}_{0}{}^{\imath}_{\alpha}(\theta) - (-1)^{\varepsilon_{
\alpha}\varepsilon_{\beta}}(\alpha \leftrightarrow \beta)\Bigr)\,,{}&\\
 {} &
{\cal R}_{0}{}^{\jmath}_{\beta},_{\imath}(\theta)
{\cal R}_{0}{}^{\imath}_{\alpha}(\theta) - (-1)^{\varepsilon_{
\alpha}\varepsilon_{\beta}}(\alpha \leftrightarrow \beta)=
 - {\cal R}_{0}{}^{\jmath}_{\gamma}(\theta)
{\cal F}_{0}{}^{\gamma}_{\beta\alpha}({\cal A}(\theta),\theta) -
S,_{\imath}(\theta)
{\cal M}_{0}{}^{\jmath\imath}_{\beta\alpha}({\cal A}(\theta),\theta)
\,,{} & \nonumber \\
{} &
{\cal F}_{0}{}^{\gamma}_{\beta\alpha}({\cal A}(\theta),\theta)\equiv
{\cal F}_{0}{}^{\gamma}_{\beta\alpha}(\theta),
{\cal M}_{0}{}^{\jmath\imath}_{\beta\alpha}({\cal A}(\theta),\theta)\equiv
{\cal M}_{0}{}^{\jmath\imath}_{\beta\alpha}(\theta) \in
C^k({\cal M}_{cl}\times\{\theta\})\,, {} &
\\
{} & \begin{array}{l|cccl}
{}& \varepsilon_P & \varepsilon_{\bar{J}} & \varepsilon &{}\\ \hline
{\cal F}_{0}{}^{\gamma}_{\alpha\beta}(\theta) & 0    &
\varepsilon_{\gamma} + \varepsilon_{\alpha}+ \varepsilon_{\beta} &
\varepsilon_{\gamma} + \varepsilon_{\alpha}+ \varepsilon_{\beta} & {}\\
{\cal M}_{0}{}^{\imath\jmath}_{\alpha\beta}(\theta) & (\varepsilon_P)_{\imath}
+ (\varepsilon_P)_{\jmath} &
(\varepsilon_{\bar{J}})_{\imath} + (\varepsilon_{\bar{J}})_{\jmath} +
\varepsilon_{\alpha} + \varepsilon_{\beta} &
\varepsilon_{\imath} + \varepsilon_{\jmath} + \varepsilon_{\alpha}+
\varepsilon_{\beta} & ,
\end{array} {} &
\\
{} & \hspace{-1.5em} {\cal F}_{0}{}^{\gamma}_{\alpha\beta}(\theta) = -
(-1)^{\varepsilon_{\alpha} \varepsilon_{\beta}}{\cal F}_{0}{}^{\gamma}_{
\beta\alpha}(\theta),\
{\cal M}_{0}{}^{\imath\jmath}_{\alpha\beta}(\theta) = -
(-1)^{\varepsilon_{\imath} \varepsilon_{\jmath}}
{\cal M}_{0}{}^{\jmath\imath}_{\alpha\beta}(\theta) = -
(-1)^{\varepsilon_{\alpha} \varepsilon_{\beta}}
{\cal M}_{0}{}^{\imath\jmath}_{\beta\alpha}(\theta).{} &
\end{eqnarray}
The explicit form of the superfunctions ${\cal F}_0{}^{\gamma}_{\alpha\beta}(
\theta)$, ${\cal M}_0{}^{\imath\jmath}_{\alpha\beta}(\theta)$
and their properties are based on the being easily proved analog of Lemma 1
in question.

\noindent
\underline{\bf Lemma 2}:

\noindent
General solution of the equation
\begin{eqnarray}
{S,}_{\imath}(\theta)y_0^{\imath}({\cal A}(\theta),\theta)=0
\end{eqnarray}
for irreducible GGTST  satisfying to condition of completeness has the form
in $C^k({\cal M}_{cl}\times\{\theta\})$
\begin{eqnarray}
{} & y_0^{\imath}(\theta)\equiv y_0^{\imath}({\cal A}(\theta),\theta) =
{\cal R}_{0}{}^{\imath}_{\gamma}({\cal A}(\theta),\theta)\Phi_0^{\gamma}({\cal A}(\theta),\theta) + {S,}_{\jmath}(\theta)E_0^{\imath\jmath}({\cal
A}(\theta),\theta)\,,{} & \\
{} & E_0^{\imath\jmath}({\cal A}(\theta),\theta)\equiv
E_0^{\imath\jmath}(\theta)=- (-1)^{\varepsilon_{\imath}\varepsilon_{
\jmath}} E_0^{\jmath\imath}(\theta),\
\Phi_0^{\gamma}({\cal A}(\theta),\theta)\equiv \Phi_0^{\gamma}(\theta)\,,
{} & \nonumber \\
{} & \begin{array}{l|cccl}
{}& \varepsilon_P & \varepsilon_{\bar{J}} & \varepsilon &{}\\\hline
\Phi_{0}^{\gamma}(\theta) & \varepsilon_P(y_0^{\imath})    &
\varepsilon_{\bar{J}}(y_0^{\imath}) + (\varepsilon_{\bar{J}})_{\imath}+
\varepsilon_{\gamma}  &
\varepsilon(y_0^{\imath}) + \varepsilon_{\imath} + \varepsilon_{\gamma} & {}\\
E_0^{\imath\jmath}(\theta) & \varepsilon_P(y_0^{\imath})+(\varepsilon_P)_{
\jmath} &
\varepsilon_{\bar{J}}(y_0^{\imath}) + (\varepsilon_{\bar{J}})_{\jmath}  &
\varepsilon(y_0^{\imath}) + \varepsilon_{\jmath} & .
\end{array} {} &
\end{eqnarray}

Identities (5.1), expression (5.10) are called in correspondence with Sec.4
and Ref.[7] by the structural equations of the 1st and 2nd orders of a
\underline{gauge algebra of GTST} respectively. Superfunctions
$S(\theta)$; ${\cal R}_0{}^{\imath}_{\alpha}(\theta)$; $
{\cal F}_0{}^{\gamma}_{\alpha\beta}(\theta)$,
${\cal M}_0{}^{\imath\jmath}_{\alpha\beta}(\theta)$   are called the
structural functions of zero; $1$st; $2$nd orders respectively. The set
of $S(\theta)$, ${\cal R}_0{}^{\imath}_{\alpha}(\theta)$, $
{\cal F}_0{}^{\gamma}_{\alpha\beta}(\theta)$,
${\cal M}_0{}^{\imath\jmath}_{\alpha\beta}(\theta)$  and so on
together with corresponding structural equations is called the gauge
algebra of GTST on $Q(S)$.

The all other concepts and remarks in the end of Sec.4 are literally
transferred onto $Q(S)$. A dependence upon ${\stackrel{\ \circ}{\cal A}}{}^{
\imath}(\theta)$ in the structural functions and equations may be only by
parametric one. The results of Sec.5 on the gauge algebra of GTST can be
obtained from gauge algebra of ordinary (not superfield on $\theta$)
irreducible gauge theory [4,6,7] by continuation of the component fields
$A^{\imath}$ to the superfields ${\cal A}^{\imath}(\theta)$ and
simultaneously by deformation on $\theta$ of the all structural functions and
equations (in the sense of their explicit dependence on $\theta$).

Gauge algebra of GTST for GThST on the whole can be efficiently described by
means of
generating equations for superfunction $S(\Gamma_{min}(\theta),\theta)$ $\in$
$C^{k}(T^{\ast}_{odd}{\cal M}_{min}$ $\times$ $\{\theta\})$, $k \le \infty$,
$T_{odd}^{\ast}{\cal M}_{min}$ = $\{(\Phi_{min}^A(\theta)$,
$\Phi_{A{}min}^{\ast}(\theta))\vert$ $\Phi_{min}^A(\theta)$ = $({\cal A}^{
\imath}(\theta), C^\alpha(\theta))$ $\in$ ${\cal M}_{min}$ = ${\cal M}_{cl}
\times {\cal M}_C$, $\Phi_{A{}min}^{\ast}(\theta)$ = $({\cal A}_{\imath}^{
\ast}(\theta)$, $C_\alpha^{\ast}(\theta))$, $A = (\imath,\alpha)\}$
which in contrast to its analog $S_{H{}min}(\Gamma_{min}(\theta))$ in
[3]  depends upon $\theta$ explicitly and is not restricted by requirement
of ordinary ghost number vanishing.

Not any GThST appears by part of a given GThGT just as not arbitrary GThGT
contains a nontrivial GThST (see corollary 2.1, 2.2 for Theorem 2 from
Ref.[1]).
However if the GThST with $S(\theta)$ is embedded into the GThGT with $Z[{
\cal A}]$ (in representing of $S_{L}(\theta)$ in the form (3.15), (3.16)) then
the corresponding gauge algebra for GThST is the gauge subalgebra in the
corresponding gauge algebra for GThGT.  Really, the vector fields
$\Gamma_{0{}\alpha}(\theta)$ (5.6) are connected with ones $\hat{\Gamma}'_{
\alpha}(\theta)$ of the type
(4.7) given and acting on $C^k({\cal M}_{cl} \times \{\theta\})$ by the
formulae
\begin{eqnarray}
{} &\left(\Gamma_{0{}\alpha}(\theta_1'){\cal F}(\theta_1')\right)\delta(
\theta_1'
-\theta_1)= \hat{\Gamma}'_{\alpha}(\theta_1){\cal F}(\theta_1'),\
{\cal F}(\theta)\in C^k({\cal M}_{cl} \times \{\theta\})\,, {} & \\
{} &
\hat{\Gamma}'_{\alpha}(\theta_1)F[{\cal A}]= \displaystyle\int \hspace{-0.4em}
d\theta\frac{\delta F[{\cal A}]}{\delta{\cal A}^{\imath}(\theta)}
\hat{\cal R}'{}_{\alpha}^{\imath}({\cal A}(\theta), \theta;{\theta}_1),\;
\hat{\cal R}'{}_{\alpha}^{\imath}({\cal A}(\theta), \theta;{\theta}_1)=
{\cal R}_{0}{}^{\imath}_{\alpha}({\cal A}(\theta), \theta)\delta(\theta-
\theta_1).{} & 
\end{eqnarray}
The structural functions $S({\cal A}(\theta), \theta)$;
${\cal R}_{0}{}^{\imath}_{\alpha}({\cal A}(\theta), \theta)$;
${\cal F}_0{}^{\gamma}_{\alpha\beta}({\cal A}(\theta), \theta)$,
${\cal M}_0{}^{\imath\jmath}_{\alpha\beta}({\cal A}(\theta),\theta)$
of zero, $1$st, $2$nd orders of the gauge algebra of
GTST are connected with corresponding ones $Z_{0}[{\cal A}]$;
$\hat{\cal R}'{}_{\alpha}^{\imath}(\theta;{\theta}')$; $
\hat{\cal F}^{\prime}{}_{\alpha\beta}^{\gamma}(\theta;{\theta}',{\theta}'_1)$,
$\hat{\cal M}'{}_{\alpha
\beta}^{\imath\jmath}(\theta,\theta_1;{\theta}', {\theta}'_1)$ of  zero,
$1$st, $2$nd orders of the gauge algebra of GTGT  by the relationships
in addition to the 2nd expression in (5.17)
\begin{eqnarray}
{} & Z_{0}[{\cal A}]= - \displaystyle\int d\theta S({\cal A}(\theta),
\theta)
\,,{} &\\ 
{} &\hat{\cal F}^{\prime}{}_{\alpha\beta}^{\gamma}({\cal A}(\theta),\theta;
\vec{\theta}_2) = (-1)^{\varepsilon_{\gamma}+\varepsilon_{\beta}}
{\cal F}_0{}^{\gamma}_{\alpha\beta}({\cal A}(\theta), \theta)
\delta(\theta-\theta_1)\delta(\theta-\theta_2)\,,{} &\\
{} &
\hat{\cal M}'{}_{\alpha\beta}^{\imath\jmath}({\cal A}(\theta'_1),
\vec{\theta}'_2;\vec{\theta}_2) = (-1)^{\varepsilon_{\jmath}+
\varepsilon_{\beta}}
{\cal M}_0{}^{\imath\jmath}_{\alpha\beta}({\cal A}(\theta'_1),\theta'_1)
\delta(\theta'_2 - \theta'_1)\times {} & \nonumber \\
{} & \frac{1}{2}[\delta(\theta'_2-\theta_1)\delta(\theta'_1-\theta_2)+
\delta(\theta'_1-\theta_1)\delta(\theta'_2-\theta_2)]\,,{} &
\end{eqnarray}
in a such way that in fulfilling of the corresponding structural equations
of the $1$st, $2$nd orders for GThST (5.1), (5.10) taking the properties
(5.11), (5.12) for ${\cal F}_0{}^{\gamma}_{\alpha\beta}(
{\cal A}(\theta), \theta)$, ${\cal M}_0{}^{\imath\jmath}_{\alpha\beta}({\cal
A}(\theta),\theta)$ into account the quantities $Z_{0}[{\cal A}]$,
$\hat{\cal R}'{}_{\alpha}^{\imath}(\theta;{\theta}')$, $
\hat{\cal F}^{\prime}{}_{\alpha\beta}^{\gamma}(\theta;{\theta}',{\theta}'_1)$
$\hat{\cal M}'{}_{\alpha\beta}^{\imath\jmath}(\theta,\theta_1;{\theta}',
{\theta}'_1)$ satisfy exactly
to the $1$st and $2$nd orders structural equations for the gauge algebra of
GTGT with $Z_{0}[{\cal A}]$ ($Z[{\cal A}]$ = $Z_{0}[{\cal A}]$ + $
\int d\theta T\bigl({\cal A}(\theta)$, ${\stackrel{\ \circ}{\cal A}}(\theta)
\bigr)$) (3.11), (4.13) with properties
(4.14), (4.15) for $\hat{\cal F}'$, $\hat{\cal M}'$. This embedding of the
gauge algebra for GThST on $Q(S)$ can be established in the all orders $k>2$
of the gauge algebra.

Derivation of the formulae (5.16), (5.19), (5.20) are based on the rules of
connection for
superfield derivatives $\displaystyle\frac{\delta\phantom{xxx}}{\delta{
\cal A}^{\imath}(\theta)}$,
$\displaystyle\frac{\partial\phantom{xxx}}{\partial{\cal A}^{\imath}(\theta)}$
obtained in [1].
\section{$\theta$-Superfield  Quantum Electrodynamics}
\setcounter{equation}{0}

As the initial GSTF model in the Lagrangian formulation consider the
superfield model of free spinor superfield of spin $\frac{1}{2}$ being by the
singular theory of special type [1]. The  model is described by Dirac
bispinor superfield
$\Psi(x,\theta)$ = $\bigl(\psi_{\gamma}(x,\theta)$, $\chi^{\dot{\gamma}}(x,
\theta)\bigr)^{T}$ = $\psi(x)$ + $\psi_1(x)\theta$
and by its Dirac
conjugate one $\overline{\Psi}(x,\theta)$ = $\Psi^{+}(x,\theta)\Gamma^0$ =
$\bigl(\overline{\chi}{}^{\beta}(x,\theta)$, $\overline{\psi}_{\dot{\beta}
}(x,\theta)\bigr)$ = $\overline{\psi}(x)$ + $\overline{\psi}_1(x)\theta$,
$\gamma, \beta$=$1,2$, $\dot{\gamma}, \dot{\beta}$=$\dot{1},\dot{2}$
being by elements of $(\frac{1}{2}, 0)$ $\bigoplus$ $(0, \frac{1}{2})$
reducible massive superfield (on $\theta$) representation $T$ of supergroup
$J$ = $\Pi(1,3)^{\uparrow}\times P$,
($\Pi(1,3)^{\uparrow}$ = $SO(1,3)^{\uparrow}$ $\ictimes$ $T(1,3)$) defined on
superspace ${\cal M}$ = ${\bf R}^{1,3} \times \tilde{P}$ = $\{(x^{\mu},
\theta)\}$, $\eta_{\mu\nu}$ = ${\rm diag}(1,-1,-1,-1)$.

Let us point out briefly the only condensed contents of index $\imath$ for
${\cal A}^{\imath}(\theta)$ $\mapsto$ $(\Psi(x,\theta), \overline{\Psi}(x,
\theta))$, the Grassmann
parities table, the transformation laws of $\Psi(x,\theta),
\overline{\Psi}(x,\theta)$ with respect to $T_{\mid P}$ representation, the
superfunction
$S_L(\theta)$ defining the GSTF model in question and Euler-Lagrange
equations (3.1) in the form of HCLF (3.17) respectively in Ref.[1] notations
\begin{eqnarray}
{} & {} &
\imath = (\gamma, \dot{\gamma}, \beta, \dot{\beta}, x),\
\begin{array}{lcccc}
{} & \psi(x) & \psi_1(x) & \Psi(x,\theta) & {}\\
\varepsilon_P & 0 & 1 & 0  & ,\ K(x,\theta) \in
\tilde{\Lambda}_{4\vert 0+1}(x^{\mu},\theta;{\bf C}), \\
\varepsilon_{\Pi} & 1 & 1 &  1 & K\in \{\Psi, \overline{\Psi}\},\\
\varepsilon & 1 & 0 & 1 &
\end{array}\\
{} & {} &
\delta\Psi(x,\theta) = \Psi'(x,\theta)
- \Psi(x,\theta) = -\mu {\stackrel{\circ}{\Psi}}(x,\theta) = -\mu
\psi_1(x)\,, \nonumber \\
{} & {} &
\delta\overline{\Psi}(x,\theta) = \overline{\Psi}'(x,\theta) -
\overline{\Psi}(x,\theta) = -\mu {\stackrel{\circ}{
\overline{\Psi}}}(x,\theta) = -\mu\overline{\psi}_1(x)\,,\\
{} & {} &
\hspace{-1em}S_L^{(1)}(\theta)\equiv
S_L\Bigl(\Psi(\theta), \overline{\Psi}(\theta),
{\stackrel{\circ}{\Psi}}(\theta), {\stackrel{\circ}{\overline{\Psi}}}
(\theta)\Bigr) = T\Bigl({\stackrel{\circ}{\Psi}}(\theta), {\stackrel{\circ}{
\overline{\Psi}}}(\theta)\Bigr) - S_{0}
\bigl(\Psi(\theta), \overline{\Psi}(\theta)\bigr)\,, \\ 
{} & {} &
\hspace{-1em}T^{(1)}(\theta) \equiv T\Bigl({\stackrel{\circ}{\Psi}}(
\theta),{\stackrel{\circ}{\overline{\Psi}}}(\theta)\Bigr) =
\displaystyle\int d^4 x  {\stackrel{\circ}{\overline{\Psi}}}
(x,\theta) {\stackrel{\circ}{\Psi}}(x,\theta) \equiv
\displaystyle\int d^4 x {\cal L}^{(1)}_{\rm kin}(x,\theta)\,, \\ 
{} & {} & \hspace{-1em}S^{(1)}_0(\theta) \equiv S_{0}\bigl(\Psi(\theta),
\overline{\Psi}(\theta)\bigr)= \displaystyle\int d^4x
\overline{\Psi}(x,\theta)\left(\imath
\Gamma^{\mu}\partial_{\mu}- m\right){\Psi}(x,\theta)
\equiv \displaystyle\int d^4x {\cal L}^{(1)}_0 (x,\theta)\,, \\ 
{} & {} &
{} \hspace{-1em}\displaystyle\frac{\delta_l Z[\Psi,\overline{\Psi}]}{\delta
\Psi(x,\theta)}  =  - \displaystyle\frac{\partial_l S^{(1)}_0(\theta)}{
\partial\Psi(x,\theta)} + \displaystyle\frac{d}{d\theta}
\frac{\partial_l T^{(1)}(\theta)
\phantom{x}}{\partial{\stackrel{\circ}{\Psi}}(x,\theta)} = -\bigl(\imath
\partial_{\mu} \overline{\Psi}(x,\theta)\Gamma^{\mu} +
m\overline{\Psi}(x,\theta)\bigr) = 0\,, \\
{} & {} &
{} \hspace{-1em}\displaystyle\frac{\delta_l
Z[\Psi,\overline{\Psi}]}{\delta \overline{\Psi}(x,\theta)} = -
\displaystyle\frac{\partial_l S^{(1)}_0(\theta)}{\partial
\overline{\Psi}(x,\theta)} +
\displaystyle\frac{d}{d\theta} \displaystyle\frac{\partial_l T^{(1)}(\theta)
\phantom{}}{\partial{\stackrel{\circ}{ \overline{\Psi}}}(x,\theta)} = -
\bigl(\imath\Gamma^{\mu}\partial_{\mu} - m\bigr){\Psi} (x,\theta) = 0\,,
\\
{} & {} &
\displaystyle\frac{\partial_l
S_0\bigl({\Psi}(\theta), \overline{\Psi}(\theta)\bigr)}{\partial
\Psi(x,\theta)\phantom{xxxxxx}}  =  \displaystyle\frac{\partial_{l,\theta,x}
{\cal L}^{(1)}_0(x,\theta)}{\partial \Psi(x,\theta) \phantom{xxxx}} -
\partial_{\nu}\displaystyle\frac{\partial_{l,\theta,x} {\cal L}^{(1)}_0(
x,\theta)}{\partial
\bigl(\partial_{\nu} \Psi(x,\theta)\bigr)\phantom{x}}\,, \\
{} & {} &
\displaystyle\frac{\partial_l
T\Bigl({\stackrel{\circ}{\Psi}}(\theta), {\stackrel{\circ}{
\overline{\Psi}}}(\theta)\Bigr)}{ \partial
{\stackrel{\circ}{\Psi}}(x,\theta)\phantom{xxxxxx}}  =
\displaystyle\frac{\partial_{l,\theta,x} {\cal L}^{(1)}_{\rm kin}(x,\theta)}{
\partial{\stackrel{\circ}{\Psi}}(x,\theta)\phantom{xxxx}} -
\partial_{\nu}\displaystyle\frac{\partial_{l, \theta,x} {\cal L}^{(1)}_{\rm
kin}(x,\theta)
}{\partial \Bigl(\partial_{\nu}{\stackrel{\circ}{\Psi}}(x,\theta) \Bigr)
\phantom{x}}\;.
\end{eqnarray}
Given model appears by nongauge one and is invariant
with respect to global $U(1)$ (phase) transformations with constant parameter
$\xi$ and elementary electric charge $e$
\begin{eqnarray}
{} & {} &
\Psi(x,\theta) \mapsto \Psi'(x,\theta) = {\rm exp}(-\imath e\xi)
\Psi(x,\theta),\
(\varepsilon_P, \varepsilon_{\Pi}, \varepsilon)\xi = (0,0,0),\ \xi\in{\bf R}
\,, \nonumber \\
{} & {} &
\overline{\Psi}(x,\theta) \mapsto \overline{\Psi}'(x,\theta)=
{\rm exp}(\imath e\xi)\overline{\Psi}(x,\theta)\,. 
\end{eqnarray}
Realizing the Yang-Mills type  gauge principle [8] let us change the
parameter onto arbitrary superfield $\xi(x,\theta)$. In this connection
Eqs.(6.6), (6.7) are changed onto $\theta$-superfield generalization of Dirac
equations in presence, at least, of external electromagnetic superfield ${\cal
A}^{\mu}(x,\theta)$ and corresponding superfunction $S_{L{}Q}^{(1)}(\theta)$
must be invariant with respect to following from (6.10) GTGT
\begin{eqnarray}
{} & {} &
{\cal A}^{\mu}(x,\theta) =  A^{\mu}(x) +  A^{\mu}_1(x)\theta \mapsto
{\cal A}'^{\mu}(x,\theta) = {\cal A}^{\mu}(x,\theta) + \partial^{\mu}\xi(x,
\theta)\,,\\
{} & {} &
C(x,\theta) =  C(x) +  C_1(x)\theta \mapsto
C'(x,\theta) = C(x,\theta) + {\stackrel{\circ}{\xi}}(x,\theta)
\,,\\
{} & {} &
\Psi(x,\theta) \mapsto \Psi'(x,\theta) = {\rm exp}(-\imath e\xi(x,\theta))
\Psi(x,\theta)\,,\\ 
{} & {} &
\overline{\Psi}(x,\theta) \mapsto \overline{\Psi}'(x,\theta)=
{\rm exp}(\imath e\xi(x,\theta))\overline{\Psi}(x,\theta)\,,\\ 
{} & {} & \hspace{-2em}
\begin{array}{lccccccc}
{} & {\cal A}^{\mu}(x,\theta) & {A}^{\mu}(x) & {A}_1^{\mu}(x) & C(x,\theta)&
C(x) & C_1(x) & {} \\
\varepsilon_P & 0 & 0 & 1 & 1 & 1 & 0 & \hspace{-0.5em},\
K(x,\theta) \in
\tilde{\Lambda}_{4\vert 0+1}(x^{\mu},\theta;{\bf R}), \\
\varepsilon_{\Pi} & 0 & 0 &  0 & 0 & 0 & 0 & K\in \{{\cal A}^{\mu}, C, \xi\}.
\\
\varepsilon & 0 & 0 & 1 & 1 & 1 &  0 & {}
\end{array} 
\end{eqnarray}
Note the $\varepsilon_P$ Grassmann parity value of superfield
${\cal A}^{\imath}(\theta)$ is not trivial in contrast to corresponding one
in Ref.[1] because of the ghost superfield $C(x,\theta)$ inclusion into
multiplet ${\cal A}^{\imath}(\theta)$ already on the initial level of the
model formulation.

Written in the infinitesimal form (3.18) with parameter $\delta\xi(x,\theta)$
the GTGT and GGTGT have the representation respectively under change of
superfield ${\cal A}^{\imath}(\theta)$ and index $\imath$ (6.1) contents
\begin{eqnarray}
{} & {} &
\delta_g{\cal A}^{\imath}(\theta) = \displaystyle\int d\theta'
\hat{\cal R}^{\imath}({\cal A}(\theta),\theta,\theta')\delta\xi(\theta')
= \displaystyle\int d\theta'd y
\hat{\cal R}^{\tilde{\imath}}({\cal A}(x,\theta),x,\theta;y,\theta')
\delta\xi(y, \theta'), \alpha = ([\xi], y)\,,\nonumber \\
{} & {} &
{\cal A}^{\imath}(\theta) = (\overline{\Psi}(x,\theta), {\Psi}(x,\theta),
{\cal A}^{\mu}(x,\theta),C(x,\theta)),\ \imath = (\gamma,\dot{\gamma},\beta,
\dot{\beta}, \mu,[C],x) = (\tilde{\imath}, x)
\,,\\
{} & {} &
\hat{\cal R}^{\tilde{\imath}}({\cal A}(x,\theta),x,\theta;y,\theta') =
\displaystyle\sum\limits_{k\geq 0} \left(\left(\displaystyle\frac{d}{d\theta}
\right)^k\delta(\theta - \theta')\right)
\hat{\cal R}_k^{\tilde{\imath}}({\cal A}(x,\theta),x,y,\theta) =
\nonumber \\
{} & {} & \hspace{2em}
\displaystyle\sum\limits_{k\geq 0} \left(\left(\displaystyle\frac{d}{d\theta}
\right)^k\delta(\theta - \theta')\right)
\hat{\cal R}_k^{\tilde{\imath}}({\cal A}(x,\theta),x,\theta)\delta(x-y)\,,
\\
{} & {} &
\hat{\cal R}_0^{\tilde{\imath}}({\cal A}(x,\theta),x,\theta) = \left\{
\begin{array}{ll}
\partial^{\mu},& \hspace{-0.3em}\imath=(\mu, x) \\
- \imath e \overline{\Psi}(x,\theta), & \hspace{-0.3em}
\imath=(\beta,\dot{\beta},x) \\
\imath e {\Psi}(x,\theta), & \hspace{-0.3em}\imath=(\gamma,\dot{\gamma},x)
\end{array}\right.\hspace{-0.7em},
\hat{\cal R}_1^{\tilde{\imath}}({\cal A}(x,\theta),x,\theta)=-1, \imath =
([C], x).
\end{eqnarray}
GGTGT (6.17), (6.18) forms the Abelian gauge algebra of GTGT in terminology of
Sec.4. To construct  zero order structural superfunction $S_{LQ}^{
(1)}(\theta)$ for given algebra let us introduce according to Ref.[8] the
prolonged covariant derivatives in $\theta$-superfield form with respect to
representation $T$ of supergroup $J$ (not Lorentz type)
\begin{eqnarray}
{\cal D}_A \equiv \partial_A - \imath e {\cal A}_{A}(x,\theta),
\partial_A=(\partial_{\mu}, \frac{d}{d\theta}),\;{\cal D}_A=({\cal D}_{\mu},
{\cal D}_{\theta}),\ {\cal A}_{A}(x,\theta)=({\cal A}_{\mu},
C)(x,\theta)\,.
\end{eqnarray}
The supercommutator of above derivatives leads to expression for superfield
${\cal A}_{A}(x,\theta)$ strength being invariant with respect to GTGT
(6.11)--(6.14)
\begin{eqnarray}
{} &
{\cal F}_{AB}(x,\theta)= \displaystyle\frac{\imath}{e}[{\cal D}_A,{\cal
D}_B]_s = \partial_A{\cal A}_{B}(x,\theta) - (-1)^{\varepsilon({\cal A}_A)
\varepsilon({\cal A}_B)}
\partial_B{\cal A}_{A}(x,\theta)\,,{} &\\
{} &
{\cal F}_{AB}(x,\theta)=
\left\|
\begin{array}{lr}
F_{\mu\nu} & F_{\mu [C]}\\
F_{[C]\nu} & F_{[C] [C]}
\end{array} \right\|(x,\theta) = \left\|
\begin{array}{cc}
\partial_{[\mu}{\cal A}_{\nu]} &
\partial_{\mu}C- {\stackrel{\ \circ}{\cal A}}_{\mu}\\
{\stackrel{\ \circ}{\cal A}}_{\nu}-\partial_{\nu}C  &
2{\stackrel{\,\circ}{C}}
\end{array} \right\|(x,\theta)=
{} &\nonumber \\
{} &
 -(-1)^{\varepsilon({\cal A}_A)\varepsilon({
\cal A}_B)}{\cal F}_{BA}(x,\theta),
(A,B)= ((\mu,[C]),(\nu,[C])),
\varepsilon({\cal A}_A)= (0\cdot\delta_{A\mu}, 1\cdot\delta_{A[C]})
\,.{} &
\end{eqnarray}
The following superfunctions being quadratic on ${\cal F}_{AB}(x,\theta)$
appear by the Poincare and gauge (with respect to GTGT) invariant objects
\begin{eqnarray}
{} & {} &
{\cal F}_{AB}(x,\theta){\cal F}^{AB}(x,\theta)= \left(F_{\mu\nu}F^{\mu\nu}
+ 2 F_{\mu [C]}F^{\mu [C]} + 4{\stackrel{\,\circ}{C}}{\stackrel{\,\circ}{C}}
\right)(x,\theta) \equiv
\nonumber \\
{} & {} & \hspace{2em}
-4\left({\cal L}^{(0)}_1(\partial_{\mu}{\cal A}_{\nu}(x,\theta))
+ {\cal L}^{(1)}_1\bigl(\partial_{\mu}C(x,\theta),
{\stackrel{\ \circ}{\cal A}}{}_{\nu}(x,\theta)\bigr) +
{\cal L}^{(2)}_1\bigl({\stackrel{\,\circ}{C}}(x,\theta)\bigr)\right)\,,
\nonumber \\
{} & {} &
{\cal L}^{(1)}_1\bigl(\partial_{\mu}C(x,\theta),{\stackrel{\ \circ}{\cal
 A}}{}_{\nu}(x,\theta)\bigr) \equiv 0;\
{\cal L}^{(2)}_1({\stackrel{\,\circ}{C}}(x,\theta)) =- \frac{d}{d\theta}\Bigl(
{C}(x,\theta){\stackrel{\,\circ}{C}}(x,\theta)\Bigr)\,;
\\
{} & {} &
\varepsilon_{ABCD}{\cal F}^{AB}(x,\theta){\cal F}^{CD}(x,\theta) =\left(
\varepsilon_{\mu\nu\rho\sigma}F^{\mu\nu}F^{\rho\sigma}
+ 4\varepsilon_{\mu\nu\rho [C]}F^{\mu\nu}F^{\rho[C]} +
4\varepsilon_{\mu\nu [C][C]}F^{\mu\nu}{\stackrel{\,\circ}{C}} + \right.
\nonumber \\
{} & {} & \hspace{2em}
\left. 4 \varepsilon_{\mu [C]\nu [C]}F^{\mu [C]}F^{\nu[C]} +
8\varepsilon_{\mu [C][C][C]}F^{\mu [C]}{\stackrel{\,\circ}{C}}\right)(x,
\theta)
- 4\varepsilon_{[C][C][C][C]}{\cal L}^{(2)}_1\bigl({\stackrel{\,\circ}{C}}
(x,\theta)\bigr)\,,\\
{} & {} & \hspace{-1.5em}\varepsilon_{ABCD}\hspace{-0.1em} =\hspace{-0.1em}
-(-1)^{\varepsilon({\cal A}_A)
\varepsilon({\cal A}_B)} \varepsilon_{BACD}\hspace{-0.1em}=\hspace{-0.1em}
-(-1)^{\varepsilon({\cal A}_C)\varepsilon({\cal A}_B)}\varepsilon_{ACBD}
\hspace{-0.1em}=\hspace{-0.1em}
-(-1)^{\varepsilon({\cal A}_C)\varepsilon({\cal A}_D)}\varepsilon_{ABDC}
\,.
\end{eqnarray}
Choosing the elements of superantisymmetric constant tensor
$\varepsilon_{ABCD}$ in the form being
compatible with even values of its ($\varepsilon_{ABCD}$) $\varepsilon_{P}$,
$\varepsilon_{\Pi}$, $\varepsilon$ gradings and with properties (6.24)
\begin{eqnarray}
\varepsilon_{0123}=
\varepsilon_{[C][C][C][C]}=1,\;\varepsilon_{\mu\nu\rho [C]}=
\varepsilon_{\mu [C][C][C]}=0,\;
\varepsilon_{\mu [C]\nu [C]}=-\varepsilon_{\mu\nu [C][C]}=
\varepsilon^{(1)}_{\mu\nu}= - \varepsilon^{(1)}_{\nu\mu}\,,
\end{eqnarray}
we obtain for (6.23) the result
\begin{eqnarray}
\Bigr(\varepsilon_{ABCD}{\cal F}^{AB}{\cal F}^{CD}\Bigl)(x,\theta) =
\Bigl(\varepsilon_{\mu\nu\rho\sigma}F^{\mu\nu}F^{\rho\sigma} -
4\varepsilon^{(1)}_{\mu\nu}\Bigl(F^{\mu\nu}
{\stackrel{\,\circ}{C}}  - F^{\mu [C]}F^{\nu [C]}\Bigr) +
4{\stackrel{\,\circ}{C}}{}^2\Bigr)(x,\theta)\,. 
\end{eqnarray}
In the first place, note the superfunction ${\cal L}_{1}^{(2)}\bigl({\stackrel{\,\circ}{
C}}(x,\theta)\bigr)$ is the self-dual one and with accuracy up to total
derivatives with respect to $x^{\mu}$, $\theta$  the sum of the
$2$nd and $3$rd summands in (6.26) is reduced to the form
\begin{eqnarray}
4\varepsilon^{(1)}_{\mu\nu}\left(F^{\nu\mu}
{\stackrel{\,\circ}{C}}+ F^{\mu [C]}F^{\nu [C]}\right)(x,\theta) =
4\varepsilon^{(1)}_{\mu\nu}\left({\stackrel{\ \circ}{\cal A}}{}^{\nu}
{\stackrel{\ \circ}{\cal A}}{}^{\mu}+2F^{\nu\mu}{\stackrel{\,\circ}{C}}\right)
(x,\theta)\,.
\end{eqnarray}
With regard of the last representation the superfunction $S_{L{}Q}^{(1)}(
\theta)$
being invariant with respect to  GTGT (6.11)--(6.14), defining the GThGT
with nontrivial inclusion of the ghost superfield $C(x,\theta)$ into
superfield (on $\theta$) quantum electrodynamics and addition of the
"$\tilde{\theta}$-term" (vacuum angle), leading by means of relationships
(6.23)--(6.27) to application in the electromagnetic duality theory
(see for instance Ref.[9]), has the resultant form
\begin{eqnarray}
{} & {} &
S_{L{}Q}^{(1)}(\theta) = S_{L{}Q}^{(1)}\bigl({\cal A}_A(\theta),
{\stackrel{\ \circ}{\cal A}}{}_A(\theta), \Psi(\theta), \overline{\Psi}(
\theta),{\stackrel{\circ}{\Psi}}(\theta), {\stackrel{\circ}{\overline{\Psi}}}(
\theta)\bigr) = \nonumber \\
{} & {} &
T_{\rm inv}\bigl({\cal D}_{\theta}\Psi(\theta),{\cal D}^{\ast}_{\theta}
\overline{\Psi}(\theta)\bigr)
- S^{(11)}\bigl(\Psi(\theta),\overline{\Psi}(\theta),{\cal
A}_{\mu}(\theta)\bigr)
- S_0^{(11)}\bigl({\cal A}_{A}(\theta),
{\stackrel{\ \circ}{\cal A}}{}_{A}(\theta)\bigr)\,;\\
{} & {} &
T_{\rm inv}(\theta) \equiv
T_{\rm inv}\bigl({\cal D}_{\theta}\Psi(\theta),{\cal D}^{\ast}_{\theta}
\overline{\Psi}(\theta)\bigr)
= \displaystyle\int d^4x{\cal L}_{\rm kin}^{(1)}\bigl({\cal D}_{\theta}
\Psi(x,\theta),{\cal D}^{\ast}_{\theta}\overline{\Psi}(x,\theta)\bigr)
= \nonumber \\
{} & {} &
\displaystyle\int d^4x\Bigl({\cal D}^{\ast}_{\theta}\overline{\Psi}\Bigr)
\Bigl({\cal D}_{\theta}\Psi\Bigr)(x,\theta),\;
{\cal D}^{\ast}_{\theta}\overline{\Psi}(x,\theta) = \left(\frac{d}{d\theta} +
\imath e C(x,\theta)\right)\overline{\Psi}(x,\theta)
\,;\\
{} & {} &
S^{(11)}(\theta)\equiv S^{(11)}\bigl(\Psi(\theta),\overline{\Psi}(\theta),
{\cal A}_{\mu}(\theta)\bigr) = \displaystyle\int d^4x
{\cal L}^{(1)}_0\bigl(\Psi(x,
\theta), \overline{\Psi}(x,\theta), {\cal D}_{\mu}\Psi(x,\theta)\bigr)
\,,\nonumber \\
{} & {} &
{\cal L}^{(1)}_0\bigl(\Psi(x,
\theta), \overline{\Psi}(x,\theta), {\cal D}_{\mu}\Psi(x,\theta)\bigr)=
\left(\overline{\Psi}\left(\imath
\Gamma^{\mu}{\cal D}_{\mu}- m\right){\Psi}\right)(x,\theta)\,;\\
{} & {} &
S^{(11)}_0(\theta)\equiv S^{(11)}_0\bigl({\cal A}_{A}(\theta),
{\stackrel{\ \circ}{\cal A}}{}_{A}(\theta)\bigr) =
\displaystyle\int d^4x {\cal L}^{(11)}_0\bigl({\cal A}_{A}(x,\theta),
\partial_B{\cal A}_{A}(x,\theta)\bigr)\,,\nonumber \\
{} & {} & \hspace{-0.5em}
{\cal L}^{(11)}_0(x,\theta)\hspace{-0.1em} = \hspace{-0.1em}\Bigl(
{\cal L}_{\tilde{\theta}} -
\frac{1}{4}{\cal F}_{AB}{\cal F}^{AB}\Bigr)(x,\theta) \hspace{-0.1em} =
\hspace{-0.1em} {\cal L}_{\tilde{\theta}}(x,\theta)
\hspace{-0.1em} + \hspace{-0.1em}
{\cal L}^{(0)}_1(\partial_{\mu}{\cal A}_{\nu}(x,\theta))\hspace{-0.1em} +
\hspace{-0.1em}{\cal L}^{(2)}_1({\stackrel{\,\circ}{C}}(x,\theta)),\\
{} & {} &
{\cal L}_{\tilde{\theta}}(x,\theta)= - \displaystyle\frac{\tilde{\theta}e^2}{
32\pi^2}\varepsilon_{ABCD}{\cal F}^{AB}(x,\theta){\cal F}^{CD}(x,\theta) =
\nonumber \\
{} & {} & \hspace{-1em}
- \displaystyle\frac{\tilde{\theta}e^2}{32\pi^2}\left(
\varepsilon_{\mu\nu\rho\sigma}F^{\mu\nu}(x,\theta)F^{\rho\sigma}(x,\theta)
+ 4\varepsilon^{(1)}_{\mu\nu}\bigl({\stackrel{\ \circ}{\cal A}}{}^{\nu}
{\stackrel{\ \circ}{\cal A}}{}^{\mu}- 2F^{\mu\nu}{\stackrel{\,\circ}{C}}
\bigr)(x,\theta) -
4{\cal L}^{(2)}_1\bigl({\stackrel{\,\circ}{C}}(x,\theta)\bigr)\right).
\end{eqnarray}
Superfunctions  $T_{\rm inv}(\theta)$, $S^{(11)}(\theta)$,
$S_{0}^{(11)}(\theta)$ in (6.28) are invariant  with respect to GTGT.
Euler-Lagrange equations (3.1) for
$Z^{(1)}[\Psi,\overline{\Psi}, {\cal A}_A]$ = $\int d\theta
S_{L{}Q}^{(1)}(\theta)$ read as follows
\begin{eqnarray}
{} & {} &
{} \hspace{-1em}\displaystyle\frac{\delta_l Z^{(1)}\phantom{xx}}{\delta
\Psi(x,\theta)} \hspace{-0.1em} = \hspace{-0.1em}
\displaystyle\frac{\partial_l S_{L{}Q}^{(1)}(\theta)}{
\partial\Psi(x,\theta)\phantom{}} + \displaystyle\frac{d}{d\theta}
\frac{\partial_l T_{\rm inv}(\theta)
\phantom{}}{\partial{\stackrel{\circ}{\Psi}}(x,\theta)}\hspace{-0.1em} =
\hspace{-0.1em} -\bigl(\imath
{\cal D}^{\ast}_{\mu}\overline{\Psi}\Gamma^{\mu} +
m\overline{\Psi}\bigr)(x,\theta)  + \imath e ({\stackrel{\,\circ}{C}}
\overline{\Psi})(x,\theta)= 0\,, \\
{} & {} &
{} \hspace{-1em}\displaystyle\frac{\delta_l
Z^{(1)}\phantom{xx}}{\delta \overline{\Psi}(x,\theta)} =
\displaystyle\frac{\partial_l S_{L{}Q}^{(1)}(\theta)}{
\partial \overline{\Psi}(x,\theta)\phantom{}} +
\displaystyle\frac{d}{d\theta} \displaystyle\frac{\partial_l T_{\rm
inv}(\theta)\phantom{}}{\partial{\stackrel{\circ}{\overline{\Psi}}}(x,
\theta)} =
- \Bigl(\bigl(\imath\Gamma^{\mu}{\cal D}_{\mu} - m\bigr) +
\imath e {\stackrel{\,\circ}{C}}(x,\theta)\Bigr)\Psi(x,\theta) = 0\,,\\
{} & {} &
{} \hspace{-1em}\displaystyle\frac{\delta_l
Z^{(1)}\phantom{xx}}{\delta {\cal A}_{\mu}(x,\theta)} =
-\displaystyle\frac{\partial_l \bigl(S^{(11)}(\theta)+S^{(11)}_0(\theta)
\bigr)}{\partial {\cal A}_{\mu}(x,\theta)\phantom{xxxxxxxxx}} +
\displaystyle\frac{d}{d\theta} \displaystyle\frac{\partial_l S^{(11)}_0(
\theta)\phantom{}}{\partial{\stackrel{\ \circ}{\cal A}}{}_{\mu}(x,\theta)}=
-\Bigl(\partial_{\nu}\Bigl[F^{\nu\mu} -
\displaystyle\frac{\tilde{\theta}e^2}{2\pi^2}
\varepsilon^{(1){}\nu\mu}{\stackrel{\,\circ}{C}}\Bigr] +
 \nonumber \\
{} & {} &\hspace{-1em}
e \overline{\Psi}\Gamma^{\mu}{\Psi} \Bigr)(x,\theta) = 0,\
\displaystyle\frac{\partial_{l,\theta,x} {\cal L}_{\tilde{\theta}}(x,\theta)}{
\partial {{\cal A}}^{\nu}(x,\theta)\phantom{xx}} - \partial_{\mu}
\displaystyle\frac{\partial_{l,\theta,x}
{\cal L}_{\tilde{\theta}}(x,\theta)}{\partial \bigl(\partial_{\mu}{\cal
A}^{\nu}(x,\theta)\bigr)} = -
\displaystyle\frac{\tilde{\theta}e^2}{2\pi^2}
\varepsilon^{(1){}\nu\mu}\partial_{\nu}{\stackrel{\,\circ}{C}}(x,\theta)
\,,\\ 
{} & {} &
{} \hspace{-1em}\displaystyle\frac{\delta_l
Z^{(1)}\phantom{xx}}{\delta C(x,\theta)} =
\displaystyle\frac{\partial_l T_{\rm inv}(\theta)}{\partial C(x,\theta)
\phantom{x}} - \displaystyle\frac{d}{d\theta} \displaystyle\frac{\partial_l
S^{(11)}_0(\theta)}{\partial{\stackrel{\,\circ}{C}}(x,\theta)}
= -  \displaystyle\frac{d}{d\theta}\left(\imath e\overline{\Psi}
{\Psi}+\displaystyle\frac{\tilde{\theta}e^2}{4\pi^2}
\varepsilon^{(1)}_{\mu\nu}F^{\mu\nu}\right)(x,\theta) = 0\,, 
\end{eqnarray}
appear by DCLF and represent by themselves the 1st (2nd) order with respect to
derivatives on $\theta$ and $x^{\mu}$ nonlinear partial differential
equations (6.33), (6.34) for spinor superfields ((6.35), (6.36) for
electromagnetic and ghost superfields). In view of degeneracy of the model
(6.28) the Cauchy problem setting is not trivial in question and remains out
the paper's scope.

From (6.28--6.36) it follows, in particular, the $\theta$-superfield  free
electrodynamics is described in terms of superfield ${\cal A}_{A}(x,\theta)$
by means of the superfunctional with "$\tilde{\theta}$-term"
\begin{eqnarray}
Z_{SED}[{\cal A_A}]\equiv Z^{(1)}[\Psi,\overline{\Psi}, {\cal A}_A]_{\mid
\Psi=\overline{\Psi}=0} = - \int d\theta S^{(11)}_0\bigl({\cal A}_{A}(\theta),
{\stackrel{\ \circ}{\cal A}}{}_{A}(\theta)\bigr)
\end{eqnarray}
the such that for $C(x,\theta)=0$ and in absence of the topological summand
$\varepsilon_{\mu\nu\rho\sigma}({\cal F}^{\mu\nu}{\cal F}^{\rho
\sigma})(x,\theta)$ in (6.37) it is obtained the GThST described with accuracy
up to nonessential number multipliers in $D=4$ by means of the free massless
vector superfield ${\cal A}_{\mu}(x,\theta)$ model [1]. The model with
$Z_{SED}[{\cal A_A}]$ (6.37) itself belongs to the class of GThST as well as
it follows from (6.35), (6.36).

To construct the corresponding to Sec.5 Abelian gauge algebra of GTST
being by the gauge subalgebra of the gauge algebra of GTGT with structural
functions in (6.17), (6.18), (6.28) it is
necessary to restrict the model onto hypersurface
${\stackrel{\circ}{\Psi}}(x,\theta)$ = ${\stackrel{\circ}{\overline{\Psi}
}}(x,\theta)$ = ${\stackrel{\ \circ}{\cal A}}{}_A(x,\theta)$ = $0$ and next
to determine the structural functions of $0$ and $1$st orders in
correspondence with (6.28)--(6.32)
\begin{eqnarray}
S(\Psi(\theta),\overline{\Psi}(\theta),{\cal A}_{\mu}(\theta), C(\theta))
\equiv S(\Psi(\theta),\overline{\Psi}(\theta),{\cal A}_{\mu}(\theta)) =
S^{(11)}(\theta) + S^{(11)}_0({\cal A}_{\mu}(\theta))
\end{eqnarray}
and with ${\cal R}_0^{\tilde{\imath}}({\cal A}(x,\theta),x,\theta)$ coinciding
with $\hat{\cal R}_0^{\tilde{\imath}}({\cal A}(x,\theta),x,\theta)$
(6.18) with the exception of superfield $C(x,\theta)$ not entering in (6.38).
Then the formulae (5.17), (5.18) completely establish  the embedding of the
gauge algebra of GTST into one of GTGT for given models.

At last, in setting $\theta=0$ in (6.11)--(6.18), (6.38) or equivalently using
the special involution $\ast$ [1] for $\tilde{\theta}=0$ we obtain the
ordinary component quantum electrodynamics formulation on classical level
being described by
${\cal A}^{\mu}(x)$, $\Psi(x)$, $\overline{\Psi}(x)$ [10].

Let us note the unification possibility of the gauge $A_{\mu}(x)$ and
ghost $C(x)$ $P_0(\theta)$-component fields into uniform
$P_0(\theta)$-component $A_{A}(x)$ of the uniform superfield
${\cal A}_{A}(x,\theta)$ (6.19) in order to realize the BRST symmetry in
the superfield form
and to construct the action functional(!) by means of the gauge strength of
the form (6.20) for the Yang-Mills type theories had been considered in the
paper [11] (see the references therein). However the form of Abelian
superfield ${\cal A}_{A}(x,\theta)$ (6.19) and strength ${\cal F}_{AB
}(x,\theta)$ (6.20) have exhausted the coincidence of the superfield models
from the present paper and from Ref.[11]. Their difference is traced not
only through the whole corresponding formulae spectrum, among them
leading to construction of the actions, but is based on the functionally
distinct conceptual formulations of the models.
\section{Conclusion}

The basic Theorem announced in Ref.[1] from which it follows the many
properties of GThGTs, GThSTs in the framework of the Lagrangian formulation
for GSTF is completely proved and their consequences are studied. Nontrivial
differential-algebraic structures, i.e. the gauge algebras of GTGT, GTST have
been investigated.

The general results of the Secs.2--5 have obtained  the final
confirmation on the example of the $\theta$-superfield quantum
electrodynamics, being on the classical level by ordinary
$\theta$-superfield
spinor electrodynamics, realized on the gauge principle basis [8]
(the so-called minimal inclusion of interaction). The cases of
$\theta$-superfield  scalar or vector electrodynamics may be deduced from the
free massive complex scalar superfields $\varphi(x,\theta)$, $\varphi^{
\ast}(x,\theta)$ model and free massive complex(!) vector superfield ${\cal
A}^{\mu}(x,\theta)$ in $D=4$ model, realized in fact in the Lagrangian
formulation of
GSTF in Ref.[1], by means of the algorithm from Sec.6. All
these models representing the GThGTs with Abelian gauge algebra can be
generalized in their constructing, in an obvious way, starting from the case
of the initial
interacting $\theta$-superfield massive spinor, scalar, (complex) vector
models.

Specially note the prolongation of the
derivative with respect to odd time $\theta$ have led to necessity already on
the classical level of the nontrivial inclusion of the ghost superfield
$C(x,\theta)$ playing the same role for $\frac{d}{d\theta}$ as the
electromagnetic one ${\cal A}^{\mu}(x,\theta)$ for $\partial_{\mu}$.

\begin{center}
{\large{\bf References}}
\end{center}
\begin{enumerate}
\item A.A. Reshetnyak, General Superfield Quantization Method. I. General
Superfield Theory of Fields: Lagrangian Formalism, hep-th/0210207.
\item A.A. Reshetnyak, General Superfield Quantization Method. II. General
Superfield Theory of Fields: Hamiltonian Formalism, hep-th/0303262.
\item A.A. Reshetnyak, General Superfield Quantization Method. III.
Construction of Quantization Scheme, hep-th/0304142.
\item I.A. Batalin and G.A. Vilkovisky,
Phys. Lett. B 102 (1981) 27.
\item D.M. Gitman and I.V. Tyutin,
Izv. Vuzov SSSR, Ser. Fizika (on russian), No.5 (1983) 3.
\item I.A. Batalin and G.A. Vilkovisky, Nucl. Phys. B 234 (1984) 106.
\item I.A. Batalin and G.A. Vilkovisky, J. Math. Phys. 26 (1985) 172.
\item C.N. Yang and R. Mills, Phys. Rev. 96 (1954) 191; R. Utiyama, Phys. Rev.
101 (1956) 1597.
\item  E. Witten, Phys. Lett. B 86 (1979) 283.
\item D.M. Gitman and I.V. Tyutin,
Quantization of Fields with Constraints (Springer-Verlag, Berlin and
Heidelberg, 1990).
\item C.M. Hull, B. Spence and J.L. Vazquez-Bello, Nucl. Phys. B 348 (1991)
108.
\end{enumerate}
\end {document}